%% file: mixed-regime.tex
\documentclass[preprint]{JHEP3} % 10pt is ignored!
\usepackage{epsfig,multicol,bbm}

\usepackage{graphicx}
\title{
Finite Volume Scaling of Pseudo Nambu-Goldstone Bosons in QCD
}
\author{F.~Bernardoni$^a$,
P.~H.~Damgaard$^b$,
H.~Fukaya$^{b,c}$,
P.~Hern\'andez$^a$\\
\it $^a$ Dpto. F\'isica Te\'orica-UV and IFIC-CSIC, Edificio Institutos Investigaci\'on,Apt. 22085, E-46071 Valencia, Spain \\
\it $^b$ The Niels Bohr Institute, The Niels Bohr International Academy,
Blegdamsvej 17 DK-2100 Copenhagen {\O}
Denmark \\
\it $^c$ High Energy Accelerator Research Organization (KEK),
  Tsukuba 305-0801, Japan\\
Email: \email{fabio.bernardoni@ific.uv.es},
\email{phdamg@nbi.dk},
\email{hfukaya@nbi.dk},
\email{pilar.hernandez@ific.uv.es}
}

\preprint{IFIC/08-38\\
FTUV-08-0815}

\abstract{
We consider chiral perturbation theory in a finite volume and in a mixed
regime of quark masses. %where we have 
We take $N_l$ light quarks near the chiral 
limit, in the so-called $\epsilon$-regime, while the remaining $N_h$ 
quarks are heavier and in the standard $p$-regime.
We compute in this new mixed regime the finite-size scaling of 
the light meson correlators in the scalar, pseudoscalar,
vector and axial vector channels.
Using the replica method, we easily
extend our results to the partially quenched theory.
With the help of our results,
lattice QCD simulations with 2+1 flavors can safely investigate 
pion physics with very light up and down quark masses 
even in the region where the pion's correlation 
length overcomes the size of the space-time lattice.
}
\begin{document}

%\begin{frontmatter}


% use the corauthref command within \author for corresponding author footnotes;% use the ead command for the email address,
%\begin{flushright}
%IFIC/08-38\\
%FTUV-08-mmdd
%\end{flushright}


%\ead{phdamg@nbi.dk}
%\ead{hfukaya@riken.jp}

%\begin{keyword}
%\keywords{Chiral symmetry, lattice QCD}
%\\\PACS 12.38.Gc
%\end{keyword}

%\end{frontmatter}

%\renewcommand{\theequation}{\arabic{section}.\arabic{equation}}
\newcommand{\Tint}[1]{{\hbox{$\sum$}\!\!\!\!\!\!\int}_{\!\!\!\!#1}}
\newcommand{\D}{{\cal D}}
\newcommand{\la}[1]{\label{#1}}
\newcommand{\be}{\begin{equation}}
\newcommand{\ee}{\end{equation}}
\newcommand{\ba}{\begin{eqnarray}}
\newcommand{\ea}{\end{eqnarray}}
\newcommand{\bi}{\begin{itemize}}
\newcommand{\ei}{\end{itemize}}
\newcommand{\rmi}[1]{{\mbox{\scriptsize #1}}}
\newcommand{\nr}[1]{(\ref{#1})}
\newcommand{\tr}{{\rm Tr\,}}
\newcommand{\re}{\mathop{\rm Re}}
\newcommand{\Hc}{{\rm H.c.\ }}
\newcommand{\im}{\mathop{\rm Im}}
\newcommand{\nn}{\nonumber \\}
\newcommand{\fr}[2]{{\frac{#1}{#2}}}
\newcommand{\msbar}{\overline{\mbox{\rm MS}}}
\newcommand{\bfp}{{\bf p}}
\newcommand{\bfx}{{\bf x}}
\newcommand{\bfi}{{\bf i}}
\newcommand{\<}{\langle} %{\left\langle}
\renewcommand{\>}{\rangle}  %{\right\rangle}
\renewcommand{\vec}[1]{{\bf #1}}
\newcommand{\pint}{\int\! dp}
\newcommand{\pslash}{\slash\!\!\! p}
\newcommand{\bmu}{\bar\mu}
%% indexing
% su4
\renewcommand{\a}{l_1}    %{\alpha}
\renewcommand{\b}{l_2}    %{\beta}
\renewcommand{\c}{l_3}    %{\gamma}
\renewcommand{\d}{l_4}    %{\delta}
\newcommand{\ha}{h_1}    %{\alpha}
\newcommand{\hb}{h_2}    %{\beta}
\newcommand{\hc}{h_3}    %{\gamma}
\newcommand{\hd}{h_4}    %{\delta}

\renewcommand{\k}{k}    %{\kappa}
\renewcommand{\l}{l}    %{\lambda}
\newcommand{\ta}{\tilde r}     %{\tilde{\a}}
\newcommand{\tb}{\tilde s}     %{\tilde{\b}}
\newcommand{\tc}{\tilde u}     %{\tilde{\c}}
\newcommand{\td}{\tilde v}     %{\tilde{\d}}
\newcommand{\RR}{{\rm I\kern -.2em  R}}
\newcommand{\eq}{Eq.~}
\newcommand{\eqs}{Eqs.~}
\newcommand{\fig}{Fig.~}
\newcommand{\figs}{Figs.~}
\newcommand{\se}{Sec.~}
\newcommand{\ses}{Secs.~}
\newcommand{\quarter}{{1\over4}}

\input{sec1.tex}

\input{sec2.tex}
\input{sec3.tex}

\input{sec4.tex}
\input{sec5.tex}
\input{sec6.tex}
\input{appendix.tex}

\input{ref.tex}
\end{document}

%% file: sec1.tex
%%%%%%%%%%%%%%%%%%%%%%%%%%%%%%%%%%%%%%%%%%%%%%%%%%%
%%%%%%%%%%%%%%%%%%%%%%%%%%%%%%%%%%%%%%%%%%%%%%%%%%%
%%%%%%%%%%%%%%%%%%%%%%%%%%%%%%%%%%%%%%%%%%%%%%%%%%%
\section{Introduction}
\label{sec:intro}
\setcounter{equation}{0}
%%%%%%%%%%%%%%%%%%%%%%%%%%%%%%%%%%%%%%%%%%%%%%%%%%%
%%%%%%%%%%%%%%%%%%%%%%%%%%%%%%%%%%%%%%%%%%%%%%%%%%%
%%%%%%%%%%%%%%%%%%%%%%%%%%%%%%%%%%%%%%%%%%%%%%%%%%%

It is now becoming feasible to simulate Quantum 
Chromodynamics (QCD) near the chiral limit in lattice
gauge theory.
Deeper theoretical understanding of chiral symmetry on
the lattice,
development of very efficient algorithms,
and a constant progress in the computational resources
have allowed a reduction of the dynamical quark masses %$m$
almost to the physical point of the $u$ and $d$ quarks.
As the quark masses approach zero, however, 
one has to be increasingly careful about finite-volume effects
since the correlation lengths of the pions,
the pseudo--Nambu-Goldstone bosons, diverge in that limit.

Such an infrared effect due to finite volume
can be systematically treated within the framework of
the low-energy effective theory.
Due to the mass gap between the lightest particles,
the pseudo--Nambu-Goldstone bosons that are generically
referred to as ``pions'', and the other hadrons, 
the heavier particles have an entirely different 
sensitivity
to the finite volume. The euclidean partition function
of QCD receives contributions from the full spectrum,
but if the chiral limit is taken at finite volume
these higher states give exponentially suppressed
contributions. In this way, by varying the volume, 
one can tune to as high 
accuracy as one wants by including only those degrees of
freedom that are associated with the pseudo--Nambu-Goldstone bosons.
In QCD, one is interested in a situation where two of the
quarks (the $u$ and the $d$) are extremely light, while
a third (the $s$) is closer to the QCD scale 
$\Lambda_{\mathrm{QCD}}$, but still light on the scale of
the ultraviolet cut-off $4\pi F$ (where $F$ is the pion
decay constant) in the effective low-energy theory.

It is therefore important to investigate
how the finite volume can affect low-energy
dynamics within the pion {\em effective} theory, 
chiral perturbation theory (ChPT) 
\cite{Gasser:1987ah}-\cite{Leutwyler:1992yt}.
An extreme finite-volume situation is reached in
the so-called $\epsilon$-regime where
the pion correlation length $1/m_\pi$ exceeds
the size $L$, $T$ of the 4-dimensional space-time 
volume $V=TL^3$,
%%%%%%%%%%%%%% Eq e-regime definition %%%%%%%%%%%%%%%%%
\begin{eqnarray}
\frac{1}{\Lambda_{\mathrm{QCD}}}~~ \ll~~ L,T~~ \ll~~ \frac{1}{m_\pi} ~.
\end{eqnarray}
The lower bound is to ensure validity of the effective chiral
theory, the coupling constants of which 
are the same as those at infinite volume.

A systematic expansion exists in
the $\epsilon$-regime, where all zero-momentum modes
of the pseudo--Nambu-Goldstone bosons have to be treated exactly.
An appropriate power-counting in this regime is
%%%%%%%%%%%%%% Eq epsilon expansion %%%%%%%%%%%%%%%%%%%
\begin{eqnarray}
\sqrt{m_q} \sim m_\pi ~\sim~ p^2 ~\sim~ 1/L^2 ~\sim~ 1/T^2 ~\sim~ \epsilon^2,
\end{eqnarray}
in units of the cut-off of the theory. With this counting, 
the  operators in the Chiral Lagrangian have different weights than in 
the ordinary ChPT at infinite volume, known as
the $p$-expansion. There is therefore a re-ordering of the perturbative 
expansion:  in many cases, the infinite-volume chiral condensate $\Sigma$ 
and the pion decay constant $F$ (both in the massless limit), 
play a more prominent role than in the conventional
large volume regime, since next-to-leading order corrections are 
calculable in terms of these leading-order couplings alone. This opens 
up the possibility of extracting
some of these low-energy constants from lattice QCD in new ways 
\cite{Gasser:1987ah}-\cite{Leutwyler:1992yt}.
%\cite{Gasser:1987ah,Neuberger:1987zz,Hansen:1990un,Hasenfratz:1989pk,
%Leutwyler:1992yt}.
   
With non-degenerate quark masses one can define an 
$\epsilon$-regime and a $p$-regime 
for each of the now mass-split pseudo--Nambu-Goldstone bosons.
In particular, one can also consider a {\em mixed} regime
in which some pseudo--Nambu-Goldstones obey the condition for the 
$\epsilon$-regime, while others fall into the $p$-regime
\cite{Bernardoni:2007hi}.
For the latter, the counting
rules are the usual ones of chiral perturbation theory,
%%%%%%%%%%%%%% Eq p-expansion  %%%%%%%%%%%%%%%%%
\begin{eqnarray}
m_q \sim m_\pi^2 ~\sim~ p^2 ~\sim~ 1/L^2 ~\sim~ 1/T^2 ~\sim~ \epsilon^2 .
\end{eqnarray}
A typical situation could be $u$ and $d$ quarks
so light that the physical pions are in the 
$\epsilon$-regime, 
%but the space-time volume $V$ is such that the
%physical kaons are in the $p$-regime. 
but the strange quark mass is such that the
physical kaons are in the $p$-regime. 
%The mixed mesons, composed of both very light and relatively heavy quarks
%of the $p$-regime behave like those of the $p$-regime. 
Further possibilities
open up when one considers partially quenched theories.
In those cases one can imagine situations in which all
physical $u$, $d$ and $s$ quarks are in the $p$-regime,
while valence quarks corresponding to all or some of 
these are taken closer towards the chiral limit, and thus
end up in the $\epsilon$-regime. Such situations could 
perhaps be realized in the context of 
mixed-action lattice simulations where dynamical configurations
are generated with physical quarks that are in the $p$-regime
and can be well treated by, say, ordinary Wilson fermion actions.
Valence quarks, which are taken to the chiral limit, could
then be of, say, overlap type. Another situation could be
the use of overlap quarks that are all or partly 
in the $p$-regime, while also overlap valence quarks are taken
to the $\epsilon$-regime. 

It is our belief that all these possibilities must be and will be
explored in future lattice gauge theory studies. The present
paper gives an analytical formalism for studying correlation
functions of pseudo--Nambu-Goldstone bosons in that setting. 
Apart from quark masses, the two remaining limitations are the
approach to the continuum limit, and the finite volume. In the present 
framework the finite
volume $V\gg 1/\Lambda_{QCD}^{4}$ 
is used as a tunable parameter with which to extract
physical observables.  The {\em only} extrapolation needed will
thus be the one associated with taking the continuum scaling limit.
With the new analytical formulas for partially quenched correlation
functions available one can extract far more information for
a given number of lattice configurations. In addition,
with the mixed-regime
predictions we also provide here one is effectively covering the
full $SU(3)$ flavor sector of QCD at low energy. The extent to which
the $s$ quark at the physical point is light enough to provide a
good description in terms of chiral dynamics remains to be
tested in detail on the lattice. 

One important feature of the $\epsilon$-regime
is the strong sensitivity to the topology of the gauge fields, 
a direct consequence of the finite volume. A crucial ingredient
in making the $\epsilon$-regime so useful for
lattice gauge theory computations is the fact that
the zero-momentum integrals can be performed analytically
at fixed topology,
typically in terms of Bessel functions. With different
detailed predictions for each sector of topological
charge there is a wealth of analytical results that
can be used to confront numerical lattice data. 

These analytical predictions are non-perturbative in the gauge theory
coupling. Remarkably, some of the leading-order $\epsilon$-regime
results were first derived on the basis of chiral
Random Matrix Theory \cite{Shuryak:1992pi}. It is
particularly simple to derive analytical expressions for
Dirac operator eigenvalues in that formulation 
\cite{Nishigaki:1998is}, and it is well understood how to
go between the two formulations \cite{Damgaard:1998xy}. 
These features all carry over into the mixed regime.

The computation of meson correlators 
in the $\epsilon$-regime was first performed in 
\cite{Hansen:1990un} and extended to both quenched and unquenched  
QCD at fixed topology a few years ago
\cite{Damgaard:2001xr}-\cite{Damgaard:2002qe}.
%\cite{Damgaard:2001xr,Damgaard:2001js,Damgaard:2002qe}. 
The
effect of a coupling to isospin chemical potential has also
been considered \cite{Akemann:2008vp}.
Partially quenched ChPT (PQChPT) in the $\epsilon$-regime 
was done for the chiral condensate itself
in \cite{Damgaard:1999ic}. Recently,
partially quenched space-time correlation functions were computed 
in several channels (pseudoscalar, scalar, and the left-current)
of meson correlators \cite{Damgaard:2007ep, Bernardoni:2007hi}. 
Also, the first computation of three-point correlation functions relevant 
for weak decays in the $\epsilon$-regime was done in 
\cite{Hernandez:2002ds}.
Some of these studies have led to determinations of    
the leading order low-energy coefficients 
$\Sigma$ and $F$\footnote{See, $e.g.$ ref. \cite{Necco:2007pr} for a 
recent summary of results.}
at various sectors of fixed topology 
in quenched lattice simulations (see, $e.g.$,
%\cite{Edwards:1999ra, Damgaard:1999tk, 
%Hernandez:1999cu, Giusti:2002sm, Bietenholz:2003bj, 
%Giusti:2003iq, Fukaya:2005yg, Ogawa:2005jn, Damgaard:2005ys,
%Bietenholz:2006fj,  Giusti:2008fz}), 
\cite{Edwards:1999ra}-\cite{Giusti:2008fz}), 
as well as the low-energy couplings of 
the $\Delta S=1$ Hamiltonian from three-point functions 
\cite{Giusti:2006mh,Hernandez:2008ft}. Difficulties associated with 
the quenched approximation have been discussed in   
\cite{Damgaard:2001xr,Damgaard:2002qe}. 
Recently, several groups have successfully extended this
to full QCD in or close to the $\epsilon$-regime.
This has been done both on the basis of Dirac operator eigenvalues and
space-time correlation functions
%\cite{DeGrand:2006nv,Lang:2006ab,Fukaya:2007fb,Fukaya:2007yv,
%DeGrand:2007tm, Joergler:2007sh,
%Hasenfratz:2007qe, Jansen:2007rx, Fukaya:2007pn, Hasenfratz:2008ce}. 
\cite{DeGrand:2006nv}-\cite{Hasenfratz:2008ce}.

%%%%%%%%%%%%% mixed expansion

In this paper, we present the results for various meson correlators
in the mixed regime of ChPT where $N_l$ light quarks are in the 
$\epsilon$-regime while $N_h=N_f-N_l$ quarks remain relatively 
heavy and belong to the standard $p$-regime. We have used two different 
methods to treat this regime. The first uses the same mixed-regime 
perturbative expansion that was introduced in \cite{Bernardoni:2007hi}. 
As a check, we have also used a 
new perturbative approach which has the advantage that it provides a 
smooth interpolation between the $\epsilon$-regime and the $p$-regime. The 
expected matching between the mixed regime and the standard $\epsilon$-regime 
is trivial to check in that formalism. The two methods should agree to all
orders, and we have checked explicitly that they do agree at least up
next-to-leading order in the mixed-regime power counting.

We treat the most general non-degenerate case where the required 
non-perturbative zero-mode integrals are
performed according to Ref. \cite{Damgaard:2007ep}.
The two-point functions of the light sector, for the pseudoscalar, scalar,
axial, vector channels, are then computed.
They can be used to extract the leading low-energy constants $\Sigma$ and $F$.
Because we work at next-to-leading order, there is also explicit dependence
on some of the $L_i$'s. In principle these low-energy constants
can be determined from fits to varying quark masses in the heavier sector,
as will become clear below.
The pseudoscalar and scalar channels for the disconnected diagrams are 
also given. We easily extend our results to the partially (and fully)  
quenched theory by applying the replica method. 
One can confirm that our formulae reduce to all previously
derived limiting cases of both the degenerate $N_f$-flavor theories 
and the fully quenched theory. There are new isospin-breaking
effects when the $u$ and $d$ quark masses are split, and the
existence of these terms can be used to extract additional
information from the correlators.
The new more general expressions should be helpful for
future lattice gauge theory simulations that aim at approaching
the chiral limit.

We start in Section \ref{sec:ChPT} by reviewing the  
mixed-regime perturbative expansion of \cite{Bernardoni:2007hi}.
The results for the two-point functions at next-to-leading order are 
presented in Section \ref{sec:two-point}. 
An alternative new approach is 
also briefly described there. The calculations are  
completed in Section \ref{sec:zero-mode} by 
explicitly performing the zero-mode integrals for the full, 
the partially quenched, 
and the fully quenched theories. As a check on our results,
we note the complete agreement between our two approaches and the 
correct matching between the mixed and pure $\epsilon$ regimes is then 
also explicitly confirmed.  
In Section \ref{sec:example}, we give an explicit example 
for $N_f=2+1$ theory presenting the pseudoscalar and axial vector correlators.
Conclusions and an outlook for the 
future  are presented in Section \ref{sec:conclusions}.

%% file: sec2.tex
%%%%%%%%%%%%%%%%%%%%%%%%%%%%%%%%%%%%%%%%%%%%%%%%%%%
%%%%%%%%%%%%%%%%%%%%%%%%%%%%%%%%%%%%%%%%%%%%%%%%%%%
%%%%%%%%%%%%%%%%%%%%%%%%%%%%%%%%%%%%%%%%%%%%%%%%%%%
%\section{A new  %$\epsilon$-
%expansion in 
%chiral perturbation theory}
\section{Chiral Perturbation Theory in the mixed-regime }\label{sec:ChPT}
\setcounter{equation}{0}
%%%%%%%%%%%%%%%%%%%%%%%%%%%%%%%%%%%%%%%%%%%%%%%%%%%
%%%%%%%%%%%%%%%%%%%%%%%%%%%%%%%%%%%%%%%%%%%%%%%%%%%
%%%%%%%%%%%%%%%%%%%%%%%%%%%%%%%%%%%%%%%%%%%%%%%%%%%

In this section we review the perturbative expansion  of the chiral
Lagrangian that was introduced in \cite{Bernardoni:2007hi} to treat the 
mixed regime. 
It incorporates features of both $\epsilon$
and $p$ expansions, allowing for the simultaneous
presence of quarks with masses corresponding to these
two regimes. The existence of such a mixed 
expansion will be useful for lattice simulations at the
physical points of the three lightest quark flavors $u, d$
and $s$, or, more modestly, simulations where only
associated valence quarks are taken to that limit.

Let us consider an $N_f$-flavor theory in 
a finite volume $V=L^3 T$,
%%%%%%%%%%%%%%% Eq ChPT Lagrangian %%%%%%%%%%%%%%%
\begin{eqnarray}
\mathcal{L}
&=&
\frac{F^2}{4}{\rm Tr}
[\partial_\mu U(x)^\dagger\partial_\mu U(x) ]
-\frac{\Sigma}{2}{\rm Tr}
[\mathcal{M}^{\dagger}U(x)U_\theta
+U_\theta^\dagger U(x)^\dagger \mathcal{M}]+\cdots,
\end{eqnarray}
where $U(x)\in SU(N_f)$ and $U_\theta=\exp(i\theta/N_f) \vec{I}$.
Here $\theta$ is a QCD vacuum angle, introduced here only
in order to be able to project on fixed gauge field topology
by doing a Fourier transform in $\theta$. 
As usual, $\Sigma$ denotes the infinite-volume chiral condensate 
in the massless limit, and
$F$ is similarly the pion decay constant in the chiral limit.
Note that there are next-to-leading order terms, indicated here
by ellipses,
each of which correspond to additional low-energy constants
denoted by, in the $SU(3)$ case,  $L_i$.

For the mass matrix $\mathcal{M}=\mathrm{diag} (m_1,m_2\cdots)$, 
we consider the most general
non-degenerate case, where we have $N_l$ light quark masses
in the $\epsilon$-regime:
%%%%%%%%%%%%%%% Eq epsilon-regime mass  %%%%%%%%%%%%%%%
\begin{eqnarray}
\mathcal{M}_{l_i l_i}\equiv m_{l_i} \sim \mathcal{O}(1/V), 
\end{eqnarray}
while the other $N_h=N_f-N_l$ quarks are heavier:
%%%%%%%%%%%%%%% Eq p-regime mass  %%%%%%%%%%%%%%%
\begin{eqnarray}
\mathcal{M}_{h_i h_i}\equiv m_{h_i} \sim \mathcal{O}(1/V^{1/2}), 
\end{eqnarray}
in units of the cut-off of the theory.
Here and in the following, we put a subscript $l$ for the
light sector and $h$ for the heavier sector and denote the mass 
matrices in those sectors by:
\begin{eqnarray}
\mathcal{M}_l \equiv \vec{P_l} \mathcal{M} \vec{P_l}  \;\;\; 
\mathcal{M}_h \equiv \vec{P_h} \mathcal{M} \vec{P_h} , 
\end{eqnarray}
where $\vec{P_l}, \vec{P_h}$ are projectors on the {\it light} and 
{\it heavier} sectors respectively. The working assumption is of course 
always that chiral perturbation theory is meaningful
even for the heavier sector.

%Following the standard procedure in the $\epsilon$-regime, 
%we first factorize the zero-mode $U$ and non-zero modes $\xi(x)$, 
%%%%%%%%%%%%%%%% Eq U xi factorize  %%%%%%%%%%%%%%%
%\begin{equation}  
%\label{eq:factorize}
%U(x) ~=~ U \exp(i\sqrt{2}\xi(x)/F) ~,
%\end{equation}
%and perform the chiral expansion in $\xi(x)$.

In ref. \cite{Bernardoni:2007hi} an expansion was proposed according to the 
following counting rules:
%%%%%%%%%%%%%%% Eq counting rule 1  %%%%%%%%%%%%%%%
\begin{eqnarray}
\label{eq:counting1}
p_\mu \sim {\cal O}(\epsilon),\;\;\;
L,T \sim {\cal O}(1/\epsilon),\;\;\;
\mathcal{M}_{ll}\sim {\cal O}(\epsilon^4),\;\;\;
\mathcal{M}_{hh}\sim {\cal O}(\epsilon^2).
\label{eq:pc}
\end{eqnarray}
An inspection of the pion propagator shows that the  zero modes of the 
Nambu-Goldstone fields associated with the generators in 
$SU(N_l)$ need to be treated non-perturbatively. All the remaining zero-modes are perturbative 
when $N_l, N_h \neq 0$. Some subtleties appear however in the partially-quenched case where all the light quarks 
are quenched, that is the replica limit $N_l \rightarrow 0$ \footnote{As usual the fully-quenched case
$N_l + N_h=0$ requires the presence of the singlet to be well-defined, but as long as $N_l + N_h \neq 0$ the singlet decouples.}.  In this case, it is easy to see that the Goldstone field associated to the generator $T_\eta$, 
\begin{eqnarray}
T_\eta = \left(\begin{array}{cc} {1\over 2 N_l} \vec{I}_{l} & 0 \\
0 &  - {1\over 2 N_h} \vec{I}_{h} \end{array}\right), 
\end{eqnarray}
gets massless in the replica limit $N_l =0$.
Here $\vec{I}_{l/h}$ are the identity matrices in the light and heavier sectors.
Note that $T_\eta$ looks ill-defined when $N_l= 0$ but
keeping $N_l$ finite until the very end of the calculation,
one sees that the replica limit $N_l\to 0$ can be safely taken.

To treat all cases on the same footing, we 
%always integrate this $\eta$ modenon-perturbatively.
%its zero-mode is, in principle, non-perturbative.  
%We  
therefore consider the following parametrization 
\footnote{In reference \cite{Bernardoni:2007hi} a different parametrization 
was considered for the case when some or all of the light quarks are 
dynamical. In that case the $\eta$ zero-mode is also perturbative and can be 
included in $\xi$. It turns out that the parametrization of 
eq.~(\ref{eq:valenciaparam}) simplifies the calculations and allows one
to consider the full and partially quenched cases on the same footing. 
Therefore we consider only eq.~(\ref{factorization2}) 
in the present paper. 
We have checked that both give the same result in the full case.}: 
\begin{equation}
\label{factorization2}
U(x)=\exp\left(\frac{2i\xi(x)}{F}\right) 
\left(    
\begin{array}{cc}
U_0 & 0 \\
0   & \vec{I}_h 
\end{array}
\right)  \exp(i \eta T_\eta),
\label{eq:valenciaparam}
\end{equation}
where $U_0 \in SU(N_l)$ is a constant matrix, $\eta$ is the zero-mode of the 
Nambu-Goldstone field associated with the $T_\eta$ generator. The $\xi$ 
fields contain the non-zero modes corresponding to all Nambu-Goldstone 
fields, and also all zero modes of those degrees of freedom 
that are not treated separately. They therefore satisfy the constraints
\begin{eqnarray}
\int d^4 x~Ê {\rm Tr}[T_a \xi(x)] = \int d^4 x ~{\rm Tr}[T_\eta \xi(x)] = 0,
\label{eq:constraint} 
\end{eqnarray}
where $T_a$ is a generator of the subgroup $SU(N_l)$. Note that the zero-mode of the $T_\eta$ generator is not included in the $\xi$ field (it is projected out by the second constraint in eq.~(\ref{eq:constraint})), and included explicitely in the last term of eq.~Ê(\ref{eq:valenciaparam}). 

We are interested in computing the correlation functions in sectors of fixed topology. Following the same derivation in \cite{Bernardoni:2007hi} we rewrite 
\begin{eqnarray}
U(x) U_\theta = U(x) \left(\begin{array}{cc}
 e^{ {i\theta\over 2 N_l}} \vec{I}_l & 0 \\
0   & e^{ {i\theta\over 2 N_h}} \vec{I}_h 
\end{array}\right)
= \exp\left(\frac{2i\xi(x)}{F}\right) 
\left(    
\begin{array}{cc}
\bar{U}_0 & 0 \\
0   & e^{- i {\bar{\eta}\over N_h}} \vec{I}_h 
\end{array}
\right),
\end{eqnarray}
where we have defined
\begin{eqnarray}
\bar{\eta}\equiv {\eta-\theta\over 2}\;\;\;\;\;\bar{\theta}\equiv {\eta+\theta\over 2}
\end{eqnarray}
and  $\bar{U}_0 \in U(N_l)$ with $\det \bar{U}_0 = e^{i \bar{\theta}} \det(U_0) = e^{i \bar{\theta}}$. 
The partition functional in sectors of fixed topology can then be written as:
  \begin{eqnarray}
\mathcal{Z}_{\nu}\simeq \int 
\left[d \xi\right] \left[d \bar{\eta}\right] \int_{U(N_l)} \left[d\bar{U}_0\right] ~J(\xi)~ 
 \det (\bar{U}_0)^{\nu} Ê\exp\left(-\int d^4 x {\mathcal L}(\xi,\bar{\eta},\bar{U}_0) \right), \nonumber\\
 \end{eqnarray}
 where as in the standard $\epsilon$-regime, the projection on fixed topology results in the 
 enlargement of the zero-mode integration from $SU(N_l)$ to $U(N_l)$. 
 $J(\xi)$ is the Jacobian of the change of variables of 
eq.~(\ref{eq:valenciaparam}). According to the power-counting of eq.~(\ref{eq:pc}), it can be shown that a consistent power-counting for the fields $\xi$ is:
\begin{eqnarray}
\xi \sim {\mathcal O}(\epsilon),
\label{eq:pc-2} 
\end{eqnarray}
therefore both the Lagrangian and the Jacobian can be perturbatively expanded in powers of $\xi$. 
At next-to-leading order we find 
\cite{Hansen:1990un, Bernardoni:2007hi}:
\begin{eqnarray}
J(\xi )&=&1-\frac{4}{3F^2V}\int d^4 x  
\sum _{a \in SU(N_l) \cup  T^{\eta} } {\rm Tr}[T_a^2\xi ^2-(T_a\xi)^2](x)  
+ \mathcal{O}(\epsilon^4) . 
 \end{eqnarray}
The Lagrangian can also be obtained as an expansion in $\epsilon$:
 \begin{eqnarray}
 {\mathcal L} &=&  \mathcal{L}^{(4)} +  \mathcal{L}^{(6)} + ...,
 \end{eqnarray}
 with terms up to $\mathcal{O}(\epsilon^4)$, up to $\mathcal{O}(\epsilon^6)$, etc .
Concerning the integration over the variable $\bar{\eta}$, we can perform a saddle-point approximation following the derivation of \cite{Bernardoni:2007hi}. 
The leading-order Lagrangian is found to be:
 \begin{eqnarray}
  \mathcal{L}^{(4)} &\equiv&  {\rm Tr}\left[\partial_{\mu}\xi \partial_{\mu}\xi\right] -\frac{\Sigma}{2}{\rm Tr} 
\left[ \mathcal{M}_{l} (\bar{U}_0+\bar{U}_0^{\dagger})\right]   
+  {2 \Sigma \over F^2} {\rm Tr}\left[  \mathcal{M}_{h}  \left(\xi - {F \over 2}{\bar{\eta} \over N_h} \vec{P}_h\right)^2\right] + i {\nu \over V} \bar{\eta} . \nonumber\\\label{eq:l4}
  \end{eqnarray}
 This quadratic form implies  also a power-counting of  ${\bar{\eta}}\sim \epsilon$. According to this rule, the last term in eq.~(\ref{eq:l4}) could be treated as a perturbation. This is true as long as $\nu \sim \mathcal{O}(\epsilon^0)$, as is usually the case in the $\epsilon$-regime. However, in the partially-quenched case $N_l = 0$, the 
  distribution of topological charge is controlled by the heavy quarks only.  Indeed the $\nu$ dependence of the leading-order partition function is found to be
  \begin{eqnarray}
  Z^{LO}_\nu \propto \exp\left( -{\nu^2\over V F^2 } \sum_h { 1 \over M_{hh}^2}\right) \int_{U(N_l)} \left[d\bar{U}_0\right]   \det (\bar{U}_0)^{\nu}  \exp\left( \frac{\Sigma}{2}\rm{Tr} \left[ \mathcal{M}_{l} (\bar{U}_0+\bar{U}_0^{\dagger})\right] \right), \nonumber\\
  \end{eqnarray}
  which in the case $N_l = 0$ implies:
  \begin{eqnarray}
  \langle \nu^2 \rangle = {1 \over 2} V F^2 {1 \over \sum_h {1 \over M_{hh}^2}} \sim \epsilon^{-2},
  \end{eqnarray}
 a scaling that makes the last term in eq.~(\ref{eq:l4}) of $\mathcal{O}(\epsilon^4)$, and therefore of leading-order.  
   In order to recover the results at $\theta=0$ by averaging over topology, it is therefore necessary to keep the last term in eq.~(\ref{eq:l4}) in the leading-order Lagrangian, or equivalently assume that $\nu \sim \epsilon^{-1}$. This is not necessary however as long as  $N_l > 0$, since the distribution of topological charge in that case is controlled by the light quarks. 
     
It is straightforward to derive the propagator for the $\xi$ fields in the light or mixed sectors from eq.~(\ref{eq:l4}), validating the power-counting of eq.~(\ref{eq:pc-2}) and ensuring that the replica limit $N_l =0$ is well-defined:
   \begin{eqnarray}
\Bigl\langle \xi_{\a\b}(x) \, \xi_{\c\d}(y) \Bigr\rangle  &=& 
 \fr12 \Bigl[\delta_{\a\d} \delta_{\b\c} {\bar \Delta}(x-y,0) - 
 \delta_{\a\b} \delta_{\c\d}  
{\bar G}(x-y, 0, 0)   \Bigr]\;, \la{gen_prop_pq_1}\\
 \Bigl\langle \xi_{\a\ha}(x) \, \xi_{\hb\b}(y) \Bigr\rangle  &=& 
 \fr12 \delta_{\a\b} \delta_{\ha\hb} 
{\Delta}\left(x-y, {M_{\ha \ha}^2\over 2}\right) \la{gen_prop_pq_2}\\
  \Bigl\langle \xi_{\a\b}(x) \, \xi_{\ha\hb}(y) \Bigr\rangle &=& 
 -\fr12 
 \delta_{\a\b} \delta_{\ha\hb} 
{{\bar G}}\left(x-y,  0, M_{\ha\ha}^2\right), \la{gen_prop_pq_3}
\end{eqnarray}
while in the heavy sector there always appear the combination:
\begin{eqnarray}
&& \Bigl\langle \left(\xi_{\ha\hb}(x)-{F  \bar{\eta}  \over 2 N_h}\delta_{\ha\hb}\right)  \, \left(\xi_{\hc\hd}(y) -{F  \bar{\eta} \over 2 N_h}\delta_{\hc\hd} \right)\Bigr\rangle  = \nonumber\\
& & \fr12 \Bigl[  \delta_{\ha\hd} \delta_{\hb\hc} {\Delta}(x-y,M_{\ha\hb}^2)  
- \delta_{\ha\hb} \delta_{\hc\hd} 
\left(\bar{G}(x-y,M_{\ha\ha}^2,M_{\hc\hc}^2) 
\right.\nonumber\\&&\left.
+ G_0(M_{\ha\ha}^2,M_{\hc\hc}^2)\right)\Bigr] , %\nonumber\\
\la{gen_prop_pq_4}
 \end{eqnarray}
where 
\begin{eqnarray}
\bar{\Delta}(x, M^2) &\equiv& \frac{1}{V}\sum_{p\neq 0}
\frac{e^{ipx}}{p^2+M^2},\;\;\; \Delta(x,M^2) 
\equiv \bar{\Delta}(x,M^2) + {1 \over V M^2}\\
\bar{G}(x, M_1^2,M_2^2) &\equiv& 
\frac{1}{V}
\sum_{p\neq 0}
\frac{e^{ipx}}{(p^2+M^2_1)(p^2+M^2_2)
\left(\frac{N_l}{p^2}+\sum^{N_h}_h \frac{1}{p^2+M^2_{hh}}\right)} \\
%G_0(M_1^2, M_2^2)&\equiv& {1 \over N_h^2 V}\left( {N_h \over M_1^2} + 
%{N_h \over  M_2^2} -  \sum_h^{N_h} {1 \over M_{hh}^2}\right),
G_0(M_1^2, M_2^2)&\equiv& {2 \nu^2 \over V^2 F^2}\left( {1\over M_1^2 M_2^2}\right),
\end{eqnarray}
and 
\begin{eqnarray}
M^2_{hh'} \equiv (m_h+m_{h'}) {\Sigma \over F^2} , 
\end{eqnarray}
is the mass of the meson fields made of the heavier quarks. 
The summation $\sum_{p\neq 0}$ is taken over
the 4-momentum
%%%%%%%%%%%%%%% Eq summation %%%%%%%%%%%%%%%%%%%%
\begin{eqnarray}
p=2\pi (n_t/T, n_x/L, n_y/L, n_z/L),
\end{eqnarray}
with integers $n_i$'s.  Note that the term $G_0$ is formally of higher order if $\nu \sim \mathcal {O}(1)$. 

In this work, one encounters $\bar{G}(x,M^2_1,M^2_2)$
with $M_1=M_2=0$ only, both in full and 
partially quenched theory\footnote{ 
The fully quenched case needs
special care; it will be discussed later.}.
In the full theory, $\bar{G}(x,0,0)$ can in principle 
be rewritten
in terms of $\bar{\Delta}$'s, as one would have expected
on general grounds. In the case of 
$N_l=2$ and $N_h=1$, which is the phenomenologically most 
interesting case, for example,
%%%%%%%%%%%%%%% Eq Gabar %%%%%%%%%%%%%%%%%%%%
\begin{eqnarray}
\bar{G}(x, 0, 0) &=& 
\frac{1}{V}
\sum_{p\neq 0}
\frac{e^{ipx}}{p^4
\left(\frac{2}{p^2}+\frac{1}{p^2+M^2_{hh}}\right)}
=\frac{1}{2}\bar{\Delta}(x,0)
-\frac{1}{6}\bar{\Delta}
(x,\frac{2}{3}M^2_{hh}).
\end{eqnarray}
But we keep using the notation of $\bar{G}(x,0,0)$ 
for simplicity in both the general case with $N_l+N_h$ flavors
and the partially quenched case, where a double pole appears.

Correlation functions are obtained by inserting 
appropriate source
operators in the above partition function and
taking suitable functional derivatives \cite{Hansen:1990un}. 
The $U(N_l)$ integral
over zero modes $\bar{U}_0$ is then done exactly, while the $\xi,\bar{\eta}$ integrals
are treated perturbatively. We return to the zero-mode
integrations in Section \ref{sec:zero-mode}. Here we will be working at 
next-to-leading order in the perturbative expansion and therefore up to the
$\mathcal{L}^{(6)}$ term in the  Lagrangian contributes:
\begin{eqnarray}
  \mathcal{L}^{(6)} &=& {2 \over 3 F^2} {\rm Tr} 
\left[(\partial_\mu \xi(x) \xi(x))^2 - (\partial_\mu \xi(x))^2 \xi^2(x)
  \right]   - {2 \Sigma \over 3 F^4} 
{\rm Tr}\left[\mathcal{M}_{h} \xi^4(x) \right]   \nonumber\\
  &+& {\Sigma \over F^2} {\rm Tr}\left[ \mathcal{M}_{l}  
\left(\xi^2(x) \bar{U}_0 + {\bar U}_0^\dagger \xi^2(x)\right) \right] + 
{16 \Sigma^2 L_4\over F^4} 
{\rm Tr}\left[ \mathcal{M}_h\right]  
{\rm Tr}\left[\partial_\mu \xi(x) \partial_\mu \xi(x) \right]  \nonumber\\
 &-& 16 {\Sigma L_6\over F^4}  {\rm Tr}\left[ \mathcal{M}_h\right]  
{\rm Tr}\left[ \mathcal{M}_{l}  \left( \bar{U}_0 + 
{\bar U}_0^\dagger \right) \right] + \ldots
   \end{eqnarray}
where the ellipses indicate terms of the same order that involve only 
$\xi_{hl}$ or $\xi_{hh}$ and do not contribute to the observables where 
valence quarks are only in the light sector, as we will be considering in this paper.

\if
The above Lagrangian has divergent expressions,
$\bar{\Delta}(0,M^2)$'s 
(and similar ones with $\bar{G}$'s as well) 
of which the explicit form of the ``chiral logarithm''
in finite volume is given by 
\cite{Hasenfratz:1989pk, Damgaard:2001xr},
%%%%%%%%%%%%%% Eq Finite volume logarithms %%%%%%%%%%
\begin{eqnarray}
\bar{\Delta}(0,M^2)&=&
\frac{M^2}{16\pi^2}(\ln \Lambda^2 +\ln V^{1/2})
-\sum_n^{\infty}\frac{1}{(n-1)!}
\beta_nM^{2(n-1)}V^{(n-2)/2} .
\end{eqnarray}
The $\beta_n$'s are the well-known shape coefficients 
that depend only on the shape of the finite 
space-time volume, and $\Lambda$ is 
the cut off of the theory.
The log-divergence $\ln \Lambda^2$ can be absorbed
in the renormalization of $L_4$ and $L_6$;
%%%%%%%%%%%%%% Eq renormalization of L_4 and L_6 %%%%
\begin{eqnarray}
L_4 \to L_4 + \frac{1}{48\times 16\pi^2}\ln \Lambda^2,\;\;\;
L_6 \to L_6 + 
\frac{1}{(16\pi)^2}\left(\frac{1}{2}
+\frac{1}{N^2_f}\right)\ln \Lambda^2
%\;\;\; (N_l\neq 0),\\
%L_6 &\to& L_6 + 
%\frac{1}{(16\pi)^2}\frac{1}{2}\ln \Lambda^2\;\;\; (N_l= 0),
\end{eqnarray}
In particular, its chiral limit is given in 
a simple form without divergence,
%%%%%%%%%%%%%% Eq Finite volume logarithms %%%%%%%%%%
\begin{eqnarray}
\bar{\Delta}(0,0)&=& -\frac{\beta_1}{V^{1/2}}.
\end{eqnarray}
\fi

%%%%%%%%%%%%% Eq e-expansion of operators %%%%%%%%%%%%%%%%% 

%% Local variables:
%% TeX-master:"./mixed-regime.tex"
%% End:

%% file: sec3.tex
%%%%%%%%%%%%%%%%%%%%%%%%%%%%%%%%%%%%%%%%%%%%%%%%%%%
%%%%%%%%%%%%%%%%%%%%%%%%%%%%%%%%%%%%%%%%%%%%%%%%%%%
%%%%%%%%%%%%%%%%%%%%%%%%%%%%%%%%%%%%%%%%%%%%%%%%%%%
\section{Two-point correlation functions }
\label{sec:two-point}
\setcounter{equation}{0}
%%%%%%%%%%%%%%%%%%%%%%%%%%%%%%%%%%%%%%%%%%%%%%%%%%%
%%%%%%%%%%%%%%%%%%%%%%%%%%%%%%%%%%%%%%%%%%%%%%%%%%%
%%%%%%%%%%%%%%%%%%%%%%%%%%%%%%%%%%%%%%%%%%%%%%%%%%%
%%%%%%%%%%%%%%%%%%%%%%%%%%%%%%%%%%%%%%%%%%%%%%%%%%%
%\subsection{The external operators}
%\label{sec:operators}
%\setcounter{equation}{0}
%%%%%%%%%%%%%%%%%%%%%%%%%%%%%%%%%%%%%%%%%%%%%%%%%%%

Correlation functions are obtained by inserting 
appropriate source operators in the above partition function and
taking suitable functional derivatives \cite{Hansen:1990un}.  
Since we consider the fully non-degenerate theory,  
we have to treat all possible $N_f\times N_f$
bilinear quark operators separately.
We therefore define
%%%%%%%%%%%%%% Eq operators QCD %%%%%%%%%%%%%%%%%%%%%%
\begin{eqnarray}
\label{eq:operators}
P^{ij}(x)=i\bar{q}_i(x)\gamma_5q_j(x),\;\;\; &
S^{ij}(x)=\bar{q}_i(x)q_j(x), \nonumber\\
A_\mu^{ij}(x)=i\bar{q}_i(x)\gamma_5\gamma_\mu q_j(x),\;\;\; &
V_\mu^{ij}(x)=i\bar{q}_i(x)\gamma_\mu q_j(x).
\end{eqnarray}

The corresponding operators in ChPT are to the leading order given by
%%%%%%%%%%%%%% Eq operators ChPT %%%%%%%%%%%%%%%%%%%%%%
\begin{eqnarray}
\label{eq:operatorsChPT}
P^{ij}(x)&=&-i\frac{\Sigma}{2}  \left( [U(x) U_\theta]_{ij}-[U_\theta^\dagger U^\dagger(x)]_{ji} \right), \\
S^{ij}(x)&=&\frac{\Sigma}{2}\left([U(x) U_\theta]_{ij}+[U_\theta^\dagger U^\dagger(x)]_{ji}\right),\\
A_\mu^{ij}(x)&=&i\frac{F^2}{2}[\partial_\mu U(x)U^\dagger(x)-
\partial_\mu U^\dagger (x)U(x)]_{ij},\\
V_\mu^{ij}(x)&=&i\frac{F^2}{2}[\partial_\mu U(x)U^\dagger(x)+
\partial_\mu U^\dagger (x)U(x)]_{ij}. 
\end{eqnarray}
The conventional irreducible representations are
obtained by appropriate combinations of $i$'s and $j$'s.
The charged pion-type meson operator and the 
neutral one are, for example, given by (we simply denote 
1 for the up quark and 2 for the down quark)
%%%%%%%%%%%%%% Eq irreducible rep %%%%%%%%%%%%%%%%%%%%%%
\begin{eqnarray}
P^{\pi^\pm} (x) = \frac{1}{2}(P^{12}(x)+P^{21}(x)),
\;\;\; &\mbox{and}&\;\;\;
P^{\pi^0} (x)= \frac{1}{2}(P^{11}(x)-P^{22}(x)).
\nonumber\\
\end{eqnarray}
%The singlet can similarly be written
%%%%%%%%%%%%%% Eq operators singlet %%%%%%%%%%%%%%%%%%%%%%
%\begin{eqnarray}
%P^{s}(x)&=&\frac{1}{N_f}\sum_i^{N_f}P^{ii}(x).
%\end{eqnarray}
In the following, we use indices $v$ and $v^\prime$
in order to specify the valence sector
which in this paper is always taken to be in the $\epsilon$-regime.

\begin{figure}[t!]\centering\hspace{0 cm}
\includegraphics[width=14cm,angle=0]{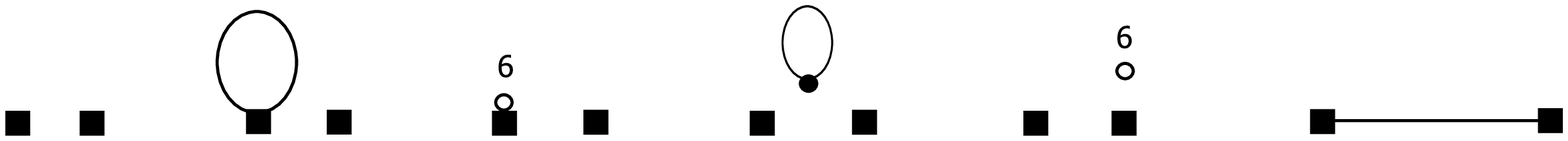} 
\vspace*{-8cm}
\caption{\label{fig:ppss} Diagrams contributing to the scalar and 
pseudoscalar connected correlators. 
The lines are $\xi$ propagators. Squares indicate 
the scalar and pseudoscalar operators. The filled dots indicate a mass 
insertion from the Lagrangian. Empty dots indicate the insertion of an 
operator coming from the NLO Lagrangian, and they are also labeled with 
the subindex of the associated coupling constant, 
$L_i$.}
\end{figure}

\begin{figure}[t!]\centering\hspace{0 cm}
\includegraphics[width=14cm,angle=0]{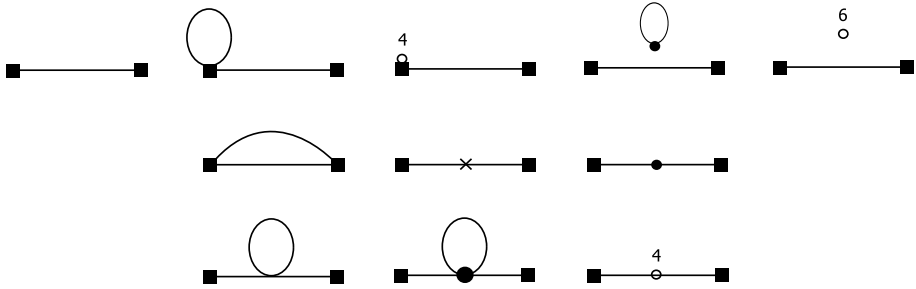}
\vspace*{-4.5cm}
\caption{\label{fig:vvaa} Diagrams contributing to the vector and axial 
correlators. The lines are $\xi$ propagators. Squares indicate the vector 
or axial vector operators. The filled dots indicate a  mass insertion. 
A cross indicate  a contribution from the Jacobian. Empty dots indicate 
the insertion of an operator coming from the NLO Lagrangian, and they are 
also labeled with the subindex of the associated coupling constant, $L_i$. }  
\end{figure}

In Figures~\ref{fig:ppss} and ~\ref{fig:vvaa} we show the Feynman diagrams 
resulting from the $\xi$ integration that contribute to the current and 
scalar propagators at next-to-leading order in the $\epsilon$-expansion. 
The scalar correlators start at ${\cal O}(\epsilon^0)$, while the first 
contribution to the currents is $\mathcal{O}(\epsilon^2)$.  
Note that disconnected diagrams contribute because they are connected 
through the zero-mode integrations. 
We also assume here that the operators are
separated from each other and the usual contact terms 
are not included.

%More unrefined non-irreducible expressions
%without 3-dimensional integrations are listed
%The results are listed in Appendix \ref{app:correlators}.

For the practical purpose of comparing to 
lattice QCD simulations, we will present results
in the non-singlet irreducible representation,
or "charged-pion" type correlation functions,
with the zero-momentum projection 
(integration over 3-dimensional space), namely,
%%%%%%%% Eq zero-mode projec definition %%%%%%%%%%
\begin{eqnarray}
%%%%% P
\mathcal{P}^c_{vv^\prime}(t)
&\equiv& \frac{1}{4}\int d^3 x 
\langle (P^{vv^\prime}(x)+P^{v^\prime v}(x))
(P^{vv^\prime}(0)+P^{v^\prime v}(0)) \rangle_{U,\xi},\\
%%%%% S
\mathcal{S}^c_{vv^\prime}(t)
&\equiv& \frac{1}{4}\int d^3 x 
\langle (S^{vv^\prime}(x)+S^{v^\prime v}(x))
(S^{vv^\prime}(0)+S^{v^\prime v}(0)) \rangle_{U,\xi},\\
%%%%% A
\mathcal{A}^c_{vv^\prime}(t)
&\equiv& \frac{1}{4}\int d^3 x 
\langle (A_0^{vv^\prime}(x)+A_0^{v^\prime v}(x))
(A_0^{vv^\prime}(0)+A_0^{v^\prime v}(0)) \rangle_{U,\xi},\\
%%%%% V
\mathcal{V}^c_{vv^\prime}(t)
&\equiv& \frac{1}{4}\int d^3 x 
\langle (V_0^{vv^\prime}(x)+V_0^{v^\prime v}(x))
(V_0^{vv^\prime}(0)+V_0^{v^\prime v}(0)) \rangle_{U,\xi},
\end{eqnarray}
where we assume $v\neq v^\prime$.

%In this section, 
We also present the "disconnected" contributions
for the scalar and pseudoscalars, 
%%%%%%% Eq discon definition %%%%%%%%%%%%
\begin{eqnarray}
%%%%%% P dis
\mathcal{P}^d_{vv^\prime}(t)
&\equiv& \int d^3 x 
\langle P^{vv}(x)P^{v^\prime v^\prime}(0)
\rangle_{U,\xi},\\
%%%%%% S dis
\mathcal{S}^d_{vv^\prime}(t)
&\equiv& \int d^3 x 
\langle S^{vv}(x)S^{v^\prime v^\prime}(0)
\rangle_{U,\xi},
\end{eqnarray}
which are useful to estimate the finite size 
contributions from the chiral fields 
to the $\eta^\prime$ meson correlators.

To simplify the $t$-dependence of our 
expressions, let us define
%%%%%%%%%% Eq h_1(t/T) definition %%%%%%%%%%%%%
\begin{eqnarray}
h_1(t/T) \equiv \frac{1}{T}\int d^3 x \bar{\Delta}(x,0)
=\frac{1}{2}\left(\frac{t}{T}-\frac{1}{2}\right)^2-\frac{1}{24},
\end{eqnarray}
and
%%%%%%%%%% Eq r(t) definition %%%%%%%%%%%%%%%%%
\begin{eqnarray}
r(t) \equiv \int d^3 x\; \bar{G}(x,0,0) .
\end{eqnarray}
Note that the latter still depends on the $N_h$ heavier quark masses.
In the appendix \ref{app:zero-momentum}, 
we list how to perform the zero-momentum projection
of various combination of $\bar{\Delta}(x, M^2)$'s.

Defining also 
\begin{eqnarray}
\label{eq:sigmatilde}
{\tilde\Sigma \over \Sigma}&\equiv&
1-\frac{1}{F^2}\left[
N_l\bar{\Delta}(0,0)+\sum^{N_h}_h \Delta(0,M^2_{hh}/2)-\bar{G}(0,0,0)
-16L_6 \sum^{N_h}_h M^2_{hh}
\right], \\
\label{eq:ftilde}
{\tilde F\over F}  &\equiv & 1
-\frac{1}{2F^2}\left[
N_l \bar{\Delta}(0,0)
+\sum^{N_h}_h \Delta(0,M^2_{hh}/2)
-8L_4\sum^{N_h}_h M^2_{hh}
\right] ,
\end{eqnarray}
and 
\begin{eqnarray}
\mu_l \equiv m_l \Sigma V, \;\;\;
\end{eqnarray}
the results for the pseudoscalar and scalar non-singlet 
(connected) correlators can be written as
\begin{eqnarray}
%%%%%%%% P non-singlet %%%%
\label{eq:results1_psci}
\mathcal{P}^c_{vv^\prime}(t)
&=&-L^3\frac{{\tilde\Sigma}^2}{4}\left[\mathcal{K}^{0(N_l)}_{-}\right]_{NLO}
+\frac{\Sigma^2}{2F^2} 
\left[ \mathcal{K}^{1(N_l)}_+T h_1\left({t\over T}\right)
-\mathcal{K}^{0(N_l)}_+ r(t) \right],\\
%%%%%%%% S non-singlet %%%%
\mathcal{S}^c_{vv^\prime}(t)
&=&\hspace{0.13in} L^3\frac{{\tilde\Sigma}^2}{4} 
\left[\mathcal{K}^{0(N_l)}_{+}\right]_{NLO}
+\frac{\Sigma^2}{2F^2} \left[\mathcal{K}^{1(N_l)}_- 
T h_1\left({t\over T}\right)
+ \mathcal{K}^{0(N_l)}_- r(t) \right],
\label{eq:results1_pscf}
\end{eqnarray}
and their disconnected correlators are given by
\begin{eqnarray}
\label{eq:results1_psdi}
\mathcal{P}^d_{vv^\prime}(t)
&=&-L^3{\tilde\Sigma}^2 \left[\mathcal{K}^{2(N_l)}_{-}\right]_{NLO}
+\frac{2\Sigma^2}{F^2}\left[\mathcal{K}^{3(N_l)}T h_1\left({t\over T}\right)
-\mathcal{K}^{2(N_l)}_+ r(t) \right],\\
\mathcal{S}^d_{vv^\prime}(t)
&=&\hspace{0.13in}L^3 {\tilde\Sigma}^2 
\left[\mathcal{K}^{2(N_l)}_{+}\right]_{NLO}
-\frac{2\Sigma^2}{F^2}\left[\mathcal{K}^{3(N_l)}T h_1\left({t\over T}\right)
- \mathcal{K}^{2(N_l)}_- r(t)\right],
\label{eq:results1_psdf}
\end{eqnarray}
The ${\mathcal K}$ functions represent the zero-mode integrals over 
$U(N_l)$ and they depend only on the light quark masses  
(to simplify the notation we denote $\bar{U}_0$ by $U$ in this section):
%%%%%%%%%% Eq U integrals %%%%%%%%%%%%%%%%%%%%%%
\begin{eqnarray}
 \label{eq:zm_psi}
\mathcal{K}^{0(N_l)}_\pm(\{\mu_l\}) &=& \frac{1}{4}
\langle (U_{vv^\prime}+ 
U_{v^\prime v} \pm 
U^\dagger_{v^\prime v} \pm U^\dagger_{v v^\prime})^2\rangle_{U(N_l)},\\
\mathcal{K}^{1(N_l)}_\pm(\{\mu_l\}) &=& 1\pm \frac{1}{2}
\langle U_{vv}U_{v^\prime v^\prime}+
U^\dagger_{vv}U^\dagger_{v^\prime v^\prime}\rangle_{U(N_l)}
\pm {1 \over 4} \langle U^2_{v v^\prime}+
U^2_{v^\prime v} + h.c.\rangle_{U(N_l)} ,\\
\mathcal{K}^{2(N_l)}_\pm(\{\mu_l\}) &=& 
\frac{1}{4}\langle (U_{vv}\pm U^\dagger_{v v})
(U_{v^\prime v^\prime}\pm U^\dagger_{v^\prime v^\prime})
\rangle_{U(N_l)},\\
\mathcal{K}^{3(N_l)}(\{\mu_l\}) &=& 
\frac{1}{4}\langle U_{vv^\prime}U_{v^\prime v}
+U^\dagger_{v^\prime v}U^\dagger_{v v^\prime})
\rangle_{U(N_l)} ,
 \label{eq:zm_psf}
\end{eqnarray}
where averages are over zero modes:
\begin{eqnarray}
\langle (...) \rangle_{U(N_l)} 
\equiv \int_{U(N_l)}~ dU (...) (\det U)^{\nu} 
e^{{\Sigma V \over 2} 
{\rm Tr}[ \vec{\mathcal M}_l  U + U^\dagger  {\mathcal M}_l^\dagger ] } .
\end{eqnarray}
The label $\left[\right]_{NLO}$ implies that the integral must be 
computed with $\tilde\Sigma$ instead of $\Sigma$.  
We will present the explicit 
results for these integrals in Section~\ref{sec:zero-mode}.

For the axial and vector (connected) current correlators we obtain:
\begin{eqnarray}
\label{eq:results1_vai}
%%%%%%%% A non-singlet %%%%
\mathcal{A}^c_{vv^\prime}(t)
&=&-\frac{{\tilde F}^2}{2T}\left[\mathcal{J}^{0(N_l)}_{+}\right]_{NLO}
+\frac{T}{2V}\left[N_lk_{00} +
\sum_h^{N_h} k^s_{00}(M^2_{hh}/2)
\right]\mathcal{J}^{0(N_l)}_-\nonumber\\
&&-\frac{\Sigma}{4}\left(\mathcal{J}^{1(N_l)}_+ + 
{2 N_l\over \Sigma V} \left(\mathcal{J}^{0(N_l)}_+ 
-\mathcal{J}^{0(N_l)}_-\right)\right)~ÊT h_1\left({t\over T}\right),\\
%%%%%%%% V non-singlet %%%%
\mathcal{V}^c_{vv^\prime}(t)
&=&-\frac{{\tilde F}^2}{2T}\left[\mathcal{J}^{0(N_l)}_{-}\right]_{NLO}
+\frac{T}{2V}\left[N_lk_{00} + 
\sum_h^{N_h} k^s_{00}(M^2_{hh}/2)\right]\mathcal{J}^{0(N_l)}_+\nonumber\\
&&-\frac{\Sigma}{4}\left(\mathcal{J}^{1(N_l)}_- + 
{2 N_l\over \Sigma V} \left(\mathcal{J}^{0(N_l)}_- 
-\mathcal{J}^{0(N_l)}_+\right)\right)~ÊT h_1\left({t\over T}\right),
\label{eq:results1_vaf}
\end{eqnarray}
where we have defined
\begin{eqnarray}
{T^2 \over V} k^s_{00}(M^2)\equiv T\frac{d}{dT}\Delta(0,M^2) \;\;\;\;
{T^2 \over V} k_{00} \equiv T\frac{d}{dT}\bar{\Delta}(0,0), 
\end{eqnarray}
and the $\mathcal{J}$ functions are given by:
%%%%%%%%%% Eq U integrals %%%%%%%%%%%%%%%%%%%%%%
\begin{eqnarray}
\label{eq:zm_vai}
\mathcal{S}^{(N_l)}_v(\{\mu_l\}) &\equiv&\frac{1}{2}\langle U_{vv}+
U^\dagger_{vv}\rangle_{U(N_l)},\\
\mathcal{J}^{0(N_l)}_\pm(\{\mu_l\}) &\equiv& 1\pm 
\frac{\langle U_{vv'}U_{vv'}^{\dagger}+
U_{vv}U_{v'v'}^{\dagger}+\mbox{h.c.} \rangle_{U(N_l)} }{2}, \\
\mathcal{J}^{1(N_l)}_\pm(\{\mu_l\}) 
&\equiv&\left((2 m_{v'}\pm m_{v}){\cal S}^{(N_l)}_{v'} \right. \nonumber\\
 &\pm & \left. \frac{\langle  U_{vv}^{\dagger}
(U{\cal M}_lU)_{v'v'}+ U_{vv'}^{\dagger}
(U{\cal M}_lU)_{vv'} +\mbox{h.c.}  
\rangle_{U(N_f)}}{2} \right) \pm  (v\leftrightarrow v^\prime)
 \label{eq:zm_vaf}
\end{eqnarray}
We stress that all the heavier mass dependence is  explicit in the 
results of eqs.~(\ref{eq:results1_psci})-(\ref{eq:results1_pscf}), 
(\ref{eq:results1_psdi})-(\ref{eq:results1_psdf}) and 
(\ref{eq:results1_vai})-(\ref{eq:results1_vaf}) since the zero-mode 
integrals involve the light sector only. We also note that these results 
agree with those obtained for the special case of the
left-handed current two-point function obtained in \cite{Bernardoni:2007hi}.

Next, we need to discuss the ultraviolet divergences of
$\Delta(0,M^2)$'s
and similar ones associated with $\bar{G}$'s. 
The explicit form in finite volume is given by 
\cite{Hasenfratz:1989pk},
%%%%%%%%%%%%%% Eq Finite volume propagators %%%%%%%%%%
\begin{eqnarray}
\Delta(0,M^2)&=&
\frac{M^2}{16\pi^2}(\ln M^2 + c_1) + g_1(M^2),
%-\sum_n^{\infty}\frac{1}{(n-1)!}
%\beta_nM^{2(n-1)}V^{(n-2)/2} .
\end{eqnarray}
where $c_1$ represents the logarithmic divergence 
which is independent of $M$ and the volume,
and $g_1$ denotes the finite volume correction 
\cite{Hasenfratz:1989pk}.
The numerical evaluation of $g_1$ is discussed in
Section \ref{sec:example}.

%which can be expressed 
%The $\beta_n$'s are 
%the well-known shape coefficients, $\beta_n$, when $ML$ is small. 
%that depend only on the geometrical shape of the finite 
%space-time volume, and $\Lambda$ is 
%the cut off of the theory.
Since $F$ and $m_i\Sigma$ are not renormalized at infinite volume,
the logarithmic divergence $c_1$ must be absorbed
in a renormalization of $L_i$'s.
The mass-independent shift in
%%%%%%%%%%%%%% Eq renormalization of L_4 and L_6 %%%%
\begin{eqnarray}
L_4 \to L_4 + \frac{1}{(16\pi)^2}c_1,\;\;\;
L_6 \to L_6 + 
\frac{1}{(16\pi)^2}\left(\frac{1}{2}
+\frac{1}{N^2_f}\right)c_1
%\;\;\; (N_l\neq 0),\\
%L_6 &\to& L_6 + 
%\frac{1}{(16\pi)^2}\frac{1}{2}\ln \Lambda^2\;\;\; (N_l= 0),
\end{eqnarray}
is enough to give finite results in the above correlators.
This shift is exactly the same as 
at infinite volume \cite{Gasser:1987ah}.

Note that the formally divergent expressions other than 
$\bar{\Delta}(0,M^2)$ (for non-zero $M$), 
%%%%%%%%%%%%%% Eq Finite volume logarithms %%%%%%%%%%
\begin{eqnarray}
\bar{\Delta}(0,0)&=& -\frac{\beta_1}{V^{1/2}},\\
k^s_{00}(M^2)&=& \sum_{\vec{q}=(p_1,p_2,p_3)}
\frac{-1}{4\sinh^2(\sqrt{|\vec{q}|^2+M^2}T/2)},
\nonumber\\
k_{00} &=& \sum_{\vec{q}=(p_1,p_2,p_3)
\neq 0}\frac{-1}{4\sinh^2(|\vec{q}|T/2)}+
\frac{1}{12}, 
\end{eqnarray}
become finite after dimensional regularization.

Finally we note that the dependence on the heavier quark masses is
as expected on general grounds (see also the discussion in 
\cite{Bernardoni:2007hi}). Indeed, up to exponentially suppressed 
finite-volume corrections in $M_{hh} L$, the correlators above  
coincide with those in the $\epsilon$-regime for $N_l$ light quarks 
as if there were no heavier quarks whatsoever. The only remnant of the
heavier quarks is seen in the modified
low-energy couplings $\Sigma$ and $F$, i.e. by the terms that depend 
on $M_{hh}$ in $\tilde\Sigma$ and $\tilde F$. This is as usual in
chiral perturbation theory.

\subsection{An alternative mixed-regime expansion}

As a check on our results, we have performed the same calculation by
means of an alternative method where the parametrization of fields
is as in the standard $\epsilon$-regime. 
The counting rule we use, however, 
is the same as the one in the standard $p$-regime for the heavy flavors. 
All zero modes in the full 
$SU(N_f)$ group are then treated non-perturbatively. Such a parametrization
has the advantage that the matching to the $\epsilon$ regime is smooth 
by construction.

The result of this alternative scheme leads to definitions of
$\tilde{\Sigma}$ and $\tilde{F}$ which are identical to eqs.
(\ref{eq:sigmatilde}) and (\ref{eq:ftilde}) except for the
replacements $\Delta \to \bar{\Delta}$ and $G \to \bar{G}$. Similarly,
all other results presented above are reproduced with the only
difference that now all zero-mode integrals are performed over the 
whole $U(N_f)$ group and therefore depend on all the quark masses, 
including the heavier ones.

In contrast to the results presented in 
\eqs.~(\ref{eq:results1_psci})-(\ref{eq:results1_psdf}) and 
(\ref{eq:results1_vai})-(\ref{eq:results1_vaf})), in this alternative
approach one can take the limit $M_{hh}\rightarrow 0$ smoothly. The results 
then coincide with those fully in the $\epsilon$-regime. Indeed,
our results in that limit agree with partially quenched
scalar and pseudoscalar correlators for non-degenerate masses that
can be found in ref. \cite{Damgaard:2007ep}. 
The left-handed current correlator can be found in \cite{Bernardoni:2007hi},
and our present results also reproduce that special case. 

The reason that the matching limit is smooth in this parametrization is 
because the zero-momentum modes of the massive mesons are resummed, 
while in the expansion of Section \ref{sec:ChPT}, they are treated 
perturbatively. 
The two results should therefore coincide when the zero-mode integrals  
of the $U(N_f)$-theory are expanded to the appropriate order in $1/\mu_h 
\sim {\cal O}(\epsilon^2)$. In the next section we will show that
this is indeed the case. This provides a rather non-trivial consistency 
check on our results, and it confirms that the expected matching between 
$\epsilon$ and mixed regimes actually holds.

%% file: sec4.tex
%%%%%%%%%%%%%%%%%%%%%%%%%%%%%%%%%%%%%%%%%%%%%%%%%%%
%%%%%%%%%%%%%%%%%%%%%%%%%%%%%%%%%%%%%%%%%%%%%%%%%%%
%%%%%%%%%%%%%%%%%%%%%%%%%%%%%%%%%%%%%%%%%%%%%%%%%%%
\section{
Non-perturbative zero-mode integrals
}
\label{sec:zero-mode}
\setcounter{equation}{0}
%%%%%%%%%%%%%%%%%%%%%%%%%%%%%%%%%%%%%%%%%%%%%%%%%%%
%%%%%%%%%%%%%%%%%%%%%%%%%%%%%%%%%%%%%%%%%%%%%%%%%%%
%%%%%%%%%%%%%%%%%%%%%%%%%%%%%%%%%%%%%%%%%%%%%%%%%%%

In this section, we complete the calculations of the correlators by
giving explicitly the zero-mode integrals defined in 
eqs.~(\ref{eq:zm_psi})-(\ref{eq:zm_psf}) and ~(\ref{eq:zm_vai})-
(\ref{eq:zm_vaf}), in the full (unquenched), 
the partially quenched, and the fully quenched theories. 
Here we present the results of general partially quenched calculations.
As is well-known, the results for the full theory can be
viewed as special cases, obtained by equating 
the valence quark masses to those of the
sea quarks. The essential ingredient is the functional
\cite{Splittorff:2002eb},
%%%%%%%%%%%%% Eq zero-mode partition function %%%%%%%%%%%%%
\begin{equation}
\label{eq:zero-mode}
\mathcal{Z}^\nu_{n,m}(\{\mu_i\})
=
\frac{\det[\mu_i^{j-1}\mathcal{J}_{\nu +j-1}(\mu_i)]_{i,j=1,\cdots n+m}}
{\prod_{j>i=1}^n(\mu_j^2-\mu_i^2)\prod_{j>i=n+1}^{n+m}(\mu_j^2-\mu_i^2)},
\end{equation}
where $\mu_i=m_i\Sigma V$.
Here $\mathcal{J}$'s are defined as
$\mathcal{J}_{\nu+j-1}(\mu_i)\equiv (-1)^{j-1} K_{\nu+j-1}(\mu_i)$ 
for $i=1,\cdots n$ and 
$\mathcal{J}_{\nu+j-1}(\mu_i)\equiv I_{\nu+j-1}(\mu_i)$ 
for $i=n+1,\cdots n+m$, 
where $I_\nu$ and $K_\nu$ are the modified Bessel functions.
$m=N_v+N_l$ denotes the number of (light) quarks, from which $N_v$ valence quarks are quenched by the $n=N_v$ bosonic quark contents. Since we are interested in mesonic two-point functions, we need $n=N_v=2$ at most.
%that can be taken to be degenerate with $\mu_i=\mu_v, i=1,...,2n$,  and $m-n$ the dynamical ones with $\{\mu_i\}_{i=n+1,...,m} = \mathcal{M}_{i-n} \Sigma V$.  The full case corresponds therefore to setting $n=0$.

%  The value of $m$ is however different in the two expansions considered in the previous section. $m=N_l$ in the first expansion and $\{\mu_i\}_{i=n+1,...,m} = \{\mu_{l_1},...,\mu_{l_{N_l}}\}$, since the zero-mode integrals are all over $U(N_l)$, while $m= N_f= N_l+N_h$ in the second expansion and $\{\mu_i\}_{i=n+1,...,m} = \{\mu_{l_1},...,\mu_{l_{N_l}},\mu_{h_1},...,\mu_{h_{N_h}}\}$, since the integrals are over the full $U(N_f)$ in that case. 

From this functional, by taking appropriate derivatives with respect to the parameters $\mu_i$, one can derive all the required integrals, both in the full as in the partially-quenched limits. 
The technical steps of our calculation have followed
those of Ref.\cite{Damgaard:2007ep} and in this section we simply show the final results.
 Details of how this can be used to compute all relevant group integrals are
presented in Appendix \ref{app:Uint}. An important relation can be derived from Ward-Takahashi identities (see Appendix \ref{app:WTI}) that holds for the full, partially-quenched and quenched cases:  
\begin{eqnarray}
{\cal J}^{1(N_l)}_{\pm} &=&2(m_{v}\pm m_{v'})({\cal S}^{(N_l)}_{v}\pm {\cal S}^{(N_l)}_{v'})\mp \frac{2 N_l}{\Sigma V}({\cal J}^{0(N_l)}_+-{\cal J}^{0(N_l)}_-). 
\end{eqnarray}

As building blocks, let us define two quantities,
%%%%%%%%%%%%%%%% Eq. building blocks
\begin{eqnarray}
\label{eq:conden}
\frac{\Sigma^{{\rm PQ}}_\nu(\mu_v,\{\mu_s\})}{\Sigma}&\equiv&
%\lim_{N\to 0} 
%\frac{1}{N}\langle\sum_v^{N}[U_0+U_0^\dagger]_{v v}
%\rangle_{U_0}
%\nonumber\\
%&=&
-\lim_{\mu_b \to \mu_v}\frac{\partial}{\partial \mu_b} 
\ln \mathcal{Z}^\nu_{1,1+N_l}(\mu_b | \mu_v, \{\mu_s\}),\\
%%%%%%%%%%%% Eq Delta Sigma definition %%%%%%%%%%%%%%%%%%%
%\begin{eqnarray}
%{\cal D}^{{\rm PQ}}_{\nu}(\mu_v,\{\mu_s\})
%&\equiv& \lim_{\mu_b\to \mu_v}\frac{
%\partial_{\mu_v} \partial_{\mu_b} 
%\mathcal{Z}^\nu_{1,1+N_l}(\mu_b|\mu_v,\{\mu_l\})}
%{\mathcal{Z}^\nu_{N_l}(\{\mu_s\})},\\
%\end{eqnarray}
%\end{eqnarray}
%and
%%%%%%%%%%% Eq D definition %%%%%%%%%%%%
%\begin{eqnarray}
D_\nu^{{\rm PQ}}(\mu_{v 1},\mu_{v 2},\{\mu_s\})
&\equiv& \hspace{2in}\nonumber\\
&&\hspace{-1.2in}
\lim_{\mu_{b 1}\to \mu_{v 1}, \mu_{b 2}\to \mu_{v 2}}
\frac{
\partial_{\mu_{v 1}}\partial_{\mu_{v 2}} 
\mathcal{Z}^\nu_{2,2+N_l}(\mu_{b 1},\mu_{b 2}|\mu_{v 1},
\mu_{v 2},\{\mu_s\})}{\mathcal{Z}^\nu_{N_l}(\{\mu_s\})}.
%\lim_{\mu_{b 1}\to \mu_{v 1}, \mu_{b 2}\to \mu_{v 2}}
%\partial_{\mu_{v 1}}\partial_{\mu_{v 2}} 
%\mathcal{Z}^\nu_{2,2+N_f}(\mu_{b 1},\mu_{b 2}|\mu_{v 1},\mu_{v 2},\{\mu_i\}).
\label{eq:Ddef}
\end{eqnarray}
%We note that in the degenerate case we have the identity 
%%%%%%%%%%% Eq D \to Delta Sigma  %%%%%%%%%%%%%
%\begin{equation}
%D_\nu^{{\rm PQ}}(\mu_v,\mu_v,\{\mu_s\})=
%- {\cal D}^{{\rm PQ}}_{\nu}(\mu_v,\{\mu_s\}) .
%\end{equation}
where the sea quark mass dependence is denoted by
$\{\mu_s\} = \{ \mu_{l_1},....,\mu_{l_{N_l}}\}$. 
%for the integrals in eqs~(\ref{eq:zm_psi})-(\ref{eq:zm_psf}) and (\ref{eq:zm_vai})-(\ref{eq:zm_vaf}) and $\{\mu_s\}=\{\mu_l, \mu_h\}= \{\mu_{l_1},...,\mu_{l_{N_l}},\mu_{h_1},...,\mu_{h_{N_h}}\}$ in our second scheme where all zero-mode integrations are done exactly.

Let us also give their analogous expressions in the unquenched theory
(we need only the case where the valence mass is equal to
one of the light sector, $m_v=m_l$ in the $\epsilon$-regime.),
%%%%%%%%%%%%%%%% Eq. building blocks full %%%%%%%%%%%
\begin{eqnarray}
\frac{\Sigma^{\rm full}_\nu (\mu_l, \{\mu_s\} )}{\Sigma}
& ~\equiv~ &
\frac{\partial}{\partial \mu_l} 
\ln \mathcal{Z}^\nu_{N_l}(\{\mu_s\})
~=~
\lim_{\mu_v\to \mu_l}
\frac{\Sigma^{{\rm PQ}}_\nu(\mu_v,\{\mu_s\})}{\Sigma}, \\
%%%%%%%%%%% Eq D definition %%%%%%%%%%%%
%\begin{eqnarray}
D_\nu^{{\rm full}}(\mu_{l 1},\mu_{l 2},\{\mu_s\})
& ~\equiv~ &
\frac{\partial_{\mu_{l 1}}\partial_{\mu_{l 2}} 
\mathcal{Z}^\nu_{N_l}(\{\mu_s\})}{\mathcal{Z}^\nu_{N_l}(\{\mu_s\})}
~=~
\lim_{\mu_{v_i} \to \mu_{li}}
D_\nu^{{\rm PQ}}(\mu_{v 1},\mu_{v 2},\{\mu_s\}),\nonumber\\
\end{eqnarray}
and the fully quenched limits,
%%%%%%%%%%%%%%%% Eq. building blocks FQ %%%%%%%%%%%
\begin{eqnarray}
\lim_{\{\mu_s\}\to \infty}
\frac{\Sigma^{{\rm PQ}}_\nu(\mu_v,\{\mu_s\})}{\Sigma}&=&
%\lim_{\mu_b \to \mu_v}
%\frac{\partial}{\partial \mu_l} 
%\ln \mathcal{Z}^\nu_{N_f}(\{\mu_s\}) \equiv 
\frac{\Sigma^{\rm FQ}_\nu (\mu_v)}{\Sigma},\\
%%%%%%%%%%% Eq D definition %%%%%%%%%%%%
%\begin{eqnarray}
\lim_{\{\mu_s\} \to \infty}
D_\nu^{{\rm PQ}}(\mu_{v 1},\mu_{v 2},\{\mu_s\})
&=& 1+\frac{\nu^2}{\mu_{v1} \mu_{v2}}.
%%%%%%%%% Check !!!!! %%%%%%%%%%%%%%%%%%
\end{eqnarray}

\subsection{Explicit results}

With the above expressions, 
one can calculate all the non-perturbative 
integrals we need to evaluate $\mathcal{J}_\pm$ {\it etc}. 
Further details can be found in Appendix \ref{app:Uint}.

\begin{enumerate}
\item{\bf Full (unquenched) theory}
\\

We start by listing the results for the full (unquenched) theory,
where the valence masses are equal to those of the sea quarks
($m_{v}=m_{l}$ and $m_{v^\prime}=m_{l^\prime}$). 
%%%%%%%%% Eq Full theory results %%%%%%%%%%%%%%%%%%%%%%%%
\begin{eqnarray}
%%% S
\mathcal{S}_l^{(N_l)}\left(\mu_l, \{\mu_s\}\right)  &=& 
\frac{\Sigma_\nu^{\mathrm{full}}(\mu_l, \{\mu_s\})}{\Sigma},\\
%%% K0
\mathcal{K}^{0(N_l)}_\pm(\mu_l, \mu_{l'},\{\mu_s\}) &=&
\frac{\pm 2}{\mu_l\mp \mu_{l^\prime}}
\left(\frac{\Sigma_\nu^{\mathrm{full}}(\mu_l, \{\mu_s\})}{\Sigma}
\mp \frac{\Sigma_\nu^{\mathrm{full}}(\mu_{l^\prime}, \{\mu_s\})}{\Sigma}
 \right),\\
%%% K1
\mathcal{K}^{1(N_l)}_\pm(\mu_l, \mu_{l'},\{\mu_s\})  &=& 1\pm \left(
D_\nu^{{\rm full}}(\mu_{l},\mu_{l^\prime},\{\mu_s\})
+\frac{\nu^2}{\mu_{l}\mu_{l^\prime}}
\right),\\
%%% K2
\mathcal{K}^{2(N_l)}_+(\mu_l, \mu_{l'},\{\mu_s\}) &=& D_\nu^{{\rm full}}(\mu_{l},\mu_{l^\prime},\{\mu_s\}),\\
\mathcal{K}^{2(N_l)}_-(\mu_l, \mu_{l'},\{\mu_s\}) &=& \frac{\nu^2}{\mu_l\mu_{l^\prime}},\\
%%% K3
\mathcal{K}^{3(N_l)}(\mu_l, \mu_{l'},\{\mu_s\})  &=& \frac{1}{\mu^2_l- \mu^2_{l^\prime}}
\left(\mu_{l^\prime}\frac{\Sigma_\nu^{\mathrm{full}}(\mu_l, \{\mu_s\})}{\Sigma}
-\mu_{l} \frac{\Sigma_\nu^{\mathrm{full}}(\mu_{l^\prime}, \{\mu_s\})}{\Sigma}
 \right),\\
 %%%% J0
 \mathcal{J}^{0(N_l)}_\pm(\mu_l, \mu_{l'},\{\mu_s\})  &=& 1\pm \left(
D_\nu^{{\rm full}}(\mu_{l},\mu_{l^\prime},\{\mu_s\})
-\frac{\nu^2}{\mu_{l}\mu_{l^\prime}}
\right),\\
%%%% J1
 \mathcal{J}^{1(N_l)}_\pm(\mu_l, \mu_{l'},\{\mu_s\})  &=& 2 (m _l \pm m_{l^\prime}) \left(\frac{\Sigma_\nu^{\mathrm{full}}(\mu_l, \{\mu_s\})}{\Sigma}
\pm \frac{\Sigma_\nu^{\mathrm{full}}(\mu_{l^\prime}, \{\mu_s\})}{\Sigma}
 \right) \nonumber\\
 &\mp & {2 N_l \over \Sigma V} \left( \mathcal{J}^{0(N_l)}_+ - \mathcal{J}^{0(N_l)}_-\right), 
 \end{eqnarray}
where $\{\mu_s\} = \{ \mu_{l_1},\mu_{l_2},\cdots \mu_{l_{N_l}}\}$. 
%for the zero-mode integrals in eqs.~(\ref{eq:zm_psi})-(\ref{eq:zm_psf}) and (\ref{eq:zm_vai})-(\ref{eq:zm_vaf}), while $N=N_f$, $\{\mu_s\} = \{\mu_l,\mu_s\}$ for those of our second scheme. 

\item{\bf Partially quenched theory ($N_l\neq 0$)}
\\

The partially quenched results where the valence masses are different
from the seq quark masses, are obtained analogously for the case $N_l\neq 0$,
%%%%%%%%% Eq PQ theory results %%%%%%%%%%%%%%%%%%%%%%%%
\begin{eqnarray}
%%% S
\mathcal{S}^{(N_l)}_v(\mu_v,\{\mu_s\})  &=& 
\frac{\Sigma_\nu^{\mathrm{PQ}}(\mu_v, \{\mu_s\})}{\Sigma},\\
%%% K0
\mathcal{K}^{0(N_l)}_\pm(\mu_v,\mu_{v'},\{\mu_s\}) &=&
\frac{\pm 2}{\mu_v\mp \mu_{v^\prime}}
\left(\frac{\Sigma_\nu^{\mathrm{PQ}}(\mu_v, \{\mu_s\})}{\Sigma}
\mp \frac{\Sigma_\nu^{\mathrm{PQ}}(\mu_{v^\prime}, \{\mu_s\})}{\Sigma}
 \right),\\
%%% K1
\mathcal{K}^{1(N_l)}_\pm(\mu_v,\mu_{v'},\{\mu_s\})  &=&1\pm \left(
D_\nu^{\mathrm{PQ}}(\mu_{v},\mu_{v^\prime},\{\mu_s\})
+\frac{\nu^2}{\mu_{v}\mu_{v^\prime}}
 \right),\\
%%% K1
\mathcal{K}^{2(N_l)}_+(\mu_v,\mu_{v'},\{\mu_s\}) &=& D_\nu^{{\rm PQ}}(\mu_{v},\mu_{v^\prime},\{\mu_s\}),\\
\mathcal{K}^{2(N_l)}_-(\mu_v,\mu_{v'},\{\mu_s\}) &=&  \frac{\nu^2}{\mu_v\mu_{v^\prime}},\\
%%% K2
\mathcal{K}^{3(N_l)}(\mu_v,\mu_{v'},\{\mu_s\}) &=& \frac{1}{\mu^2_v- \mu^2_{v^\prime}}
\left(\mu_{v^\prime}\frac{\Sigma_\nu^{\mathrm{PQ}}(\mu_v, \{\mu_s\})}{\Sigma}
-\mu_{v} \frac{\Sigma_\nu^{\mathrm{PQ}}(\mu_{v^\prime}, \{\mu_s\})}{\Sigma}
 \right),\\
 %%% J
\mathcal{J}^{0(N_l)}_\pm(\mu_v,\mu_{v'},\{\mu_s\})  &=& 1\pm \left(
D_\nu^{{\rm PQ}}(\mu_{v},\mu_{v^\prime},\{\mu_s\})
-\frac{\nu^2}{\mu_{v}\mu_{v^\prime}}
\right),\\
%%%% J1
 \mathcal{J}^{1(N_l)}_\pm(\mu_v,\mu_{v'},\{\mu_s\})  &=& 2 (m_v\pm m_{v^\prime}) \left(\frac{\Sigma_\nu^{\mathrm{PQ}}(\mu_v, \{\mu_s\})}{\Sigma}
\pm \frac{\Sigma_\nu^{\mathrm{PQ}}(\mu_{v^\prime}, \{\mu_s\})}{\Sigma}
 \right) \nonumber\\
 &\mp & {2 N_l\over \Sigma V} \left( \mathcal{J}^0_+ - \mathcal{J}^0_-\right). 
\end{eqnarray}

\item{\bf Partially quenched theory ($N_l= 0$)}
\\

In the case with $N_l=0$, one needs 
the {\it fully quenched} integral over $\bar{U}_0$:
\begin{eqnarray}
%%% S
\mathcal{S}^{(0)}_v(\mu_v)  &=& 
\frac{\Sigma_\nu^{\mathrm{FQ}}(\mu_v)}{\Sigma},\\
%%% K0
\mathcal{K}^{0(0)}_\pm(\mu_v,\mu_{v'}) &=&
\frac{\pm 2}{\mu_v\mp \mu_{v^\prime}}
\left(\frac{\Sigma_\nu^{\mathrm{FQ}}(\mu_v)}{\Sigma}
\mp \frac{\Sigma_\nu^{\mathrm{FQ}}(\mu_{v^\prime})}{\Sigma}
 \right),\\
%%% K1
\frac{\mathcal{K}^{1(0)}_+(\mu_v,\mu_{v'})}{2}  &=&
\mathcal{K}^{2(0)}_+(\mu_v,\mu_{v'})
=1
+\frac{\nu^2}{\mu_{v}\mu_{v^\prime}},\\
%%% K2
\frac{\mathcal{K}^{1(0)}_-(\mu_v,\mu_{v'})}{2}
&=&\mathcal{K}^{2(0)}_-(\mu_v,\mu_{v'}) =  \frac{\nu^2}{\mu_v\mu_{v^\prime}},\\
%%% K3
\mathcal{K}^{3(0)}(\mu_v,\mu_{v'}) &=& \frac{1}{\mu^2_v- \mu^2_{v^\prime}}
\left(\mu_{v^\prime}\frac{\Sigma_\nu^{\mathrm{FQ}}(\mu_v)}{\Sigma}
-\mu_{v} \frac{\Sigma_\nu^{\mathrm{FQ}}(\mu_{v^\prime})}{\Sigma}
 \right),\\
%%% J
\mathcal{J}^{0(0)}_+(\mu_v,\mu_{v'})  &=& 2,\;\;\;
\mathcal{J}^{0(0)}_-(\mu_v,\mu_{v'})=0.\\
%%%% J1
 \mathcal{J}^{1(0)}_\pm(\mu_v,\mu_{v'})  &=& 2 (m_v \pm m _{v^\prime}) \left(\frac{\Sigma_\nu^{\mathrm{FQ}}(\mu_v)}{\Sigma}
\pm \frac{\Sigma_\nu^{\mathrm{FQ}}(\mu_{v^\prime})}{\Sigma}
 \right). %\nonumber\\
% &\mp & {2 0\over \Sigma V} \left( \mathcal{J}^0_+ - \mathcal{J}^0_-\right). \nonumber\\
\end{eqnarray}

\item{\bf Fully quenched theory ($N_l= N_h=0$)}
\\

When $N_l=N_h=0$ or the theory is fully quenched,
the zero-mode integrals we use are the same as 
the partially quenched case with $N_l=0$ above.
But we need further to include the singlet degree 
of freedom for the non-zero modes,
with additional low-energy constants $\alpha$ and $m_0^2$
as quenched artifacts 
\cite{Bernard:1992mk,Sharpe:1992ft,Damgaard:2001js, Damgaard:2002qe}. This 
results in the modification of
$r(t)$ to 
\begin{eqnarray}
r(t) = \frac{1}{N_c}\left(-\frac{m_0^2T^3}{24}\left[
\left(\frac{t}{T}\right)^2\left(\frac{t}{T}-1\right)^2-\frac{1}{30}\right]
+\alpha Th_1(t)
\right),
\end{eqnarray}
where $N_c$ denotes the number of colors.
See $e.g.$ ref. \cite{Damgaard:2001js} for details.
\end{enumerate}

Using the unitarity condition given in ref. \cite{Damgaard:2007ep}, 
we have checked that all of the above expressions precisely reproduce
the known results for degenerate $N_l\neq 0$ flavors 
(setting $N_h=0$), and the quenched results (the $N_l=N_h=0$ limit),
obtained earlier \cite{Damgaard:2001js, Damgaard:2002qe}.

\subsection{Equivalence of the two mixed-regime expansions}

The apparent difference between results obtained in the two mixed-regime 
expansions considered in section \ref{sec:two-point} arises from the 
contribution of the zero-momentum modes of the heavy mesons. They are 
computed perturbatively in the first case and resummed in the second. 
In order for the two results 
to agree, an additional expansion in $1/\mu_h \sim \epsilon^2$ of the 
zero-mode integrals for $U(N_f)$ must of course be performed so that 
only terms at subleading order in the $\epsilon$ expansion are consistently 
kept in the correlators. Performing this expansion one finds
\footnote{We have checked these expansions in several special cases 
with a rather small number of flavors.}:
%%%%%%%%%% N_l to N_f %%%%%%%%%%%%%%%%%%
%\begin{eqnarray}
%\left.\frac{\Sigma_\nu^{\mathrm{PQ}}(\mu_v, \{\mu_l, \mu_h\})}{\Sigma}
%\right|_{N_f{\rm flavor}}
%=\left.
%\frac{\Sigma_\nu^{\mathrm{PQ}}(\tilde{\mu_v}, \{\tilde{\mu_l} \})}
%{\Sigma}\right|_{N_l {\rm flavor}}  \times 
%\left(1- \sum_h {2 \over  \mu_h}\right),
%\end{eqnarray}
%and
\begin{eqnarray}
\mathcal{J}^{0(N_f)}_\pm\left(\{\mu_l\},\{\mu_h\}\right) &=& \mathcal{J}^{0(N_l)}_\pm\left( \{\tilde{\mu}_l \}\right) \left(1- \sum_h {1 \over  \mu_h}\right) + \sum_h {1 \over \mu_h} \mathcal{J}^{0(N_l)}_\mp\left( \{\mu_l\}\right), \nonumber\\
\mathcal{J}^{1(N_f)}_\pm\left(\{\mu_l\},\{\mu_h\}\right) &=& \mathcal{J}^{1(N_l)}_\pm\left( \{\mu_l \}\right)  + {\mathcal O}\left({1\over \mu_h}\right), \nonumber\\
\mathcal{K}^{0(N_f)}_\pm\left(\{\mu_l\},\{\mu_h\}\right) &=& \mathcal{K}^{0(N_l)}_\pm\left(\{\tilde{\mu}_l\}
\right) \left(1- \sum_h {2 \over  \mu_h}\right) + \mathcal{O}\left({1 \over \mu_h}\right)^2,\nonumber\\
\mathcal{K}^{n(N_f)}_\pm\left(\{\mu_l\},\{\mu_h\}\right) &=& \mathcal{K}^{n(N_l)}_\pm\left(\{\mu_l\}\right) + \mathcal{O}\left({1 \over \mu_h}\right),\;\;\; n=1,2,3.
\end{eqnarray}
We have here denoted
\begin{eqnarray}
{\tilde\mu}_i \equiv \mu_i  \left(1 -\sum_h {1 \over \mu_h}\right). 
\end{eqnarray}
Using these expansions the results from the two different schemes agree.

As another non-trivial check in the opposite direction, 
one can also confirm that a fully perturbative approach
as in the $p$-regime,
where all of $N_f=N_l+N_h$ flavors are perturbatively treated,
is consistent with our results in an unrealistic limit 
$FL\gg 1$ while $M_\pi L <1$ kept.

%% file: sec5.tex
\section{Useful examples for 2+1 flavor theory}
\label{sec:example}
\setcounter{equation}{0}
%%%%%%%%%%%%%%%%%%%%%%%%%%%%%%%%%%%%%%%%%%%%%%%%%%%
%%%%%%%%%%%%%%%%%%%%%%%%%%%%%%%%%%%%%%%%%%%%%%%%%%%
%%%%%%%%%%%%%%%%%%%%%%%%%%%%%%%%%%%%%%%%%%%%%%%%%%%

In this section we give some explicit examples 
that are useful when
comparing with lattice QCD simulations.
Here we consider the 2+1 flavor theory where 
the up and down quark masses are degenerate,
$m_u=m_d$ and different from the strange quark mass $m_s$.
%We set the finite 4-torus with $L=2$fm and $T=4$fm.
We choose the low-energy constants to be the phenomenologically
reasonable values
% $F=90$ MeV, $\Sigma^{1/3}=250$ MeV,
%%%%%%%%%% Eq The values of LECS %%%%%%%%%%%%%%%%%%%
%\begin{eqnarray}
$F=90 {\rm MeV}$,
$\Sigma^{1/3}=250 {\rm MeV}$,
%\nonumber\\
$L^r_4(0.77{\rm GeV})=0.1\times 10^{-3}$ and
$L^r_6(0.77{\rm GeV})=0.05\times 10^{-3}$.

%\nonumber\\
%\end{eqnarray}

For the calculation of $g_1(M^2)$, we use an expansion
in the modified Bessel function \cite{Bernard:2001yj},
%%%%%%%%% Eq numerical g_1 %%%%%%%%%%%%%%%%%%%%%%%%
\begin{eqnarray}
g_1(M^2)&=&\sum^{|n_i|\leq n_{max}}_{a \neq 0}\frac{M}{4\pi^2|a|}K_1(M|a|),
\end{eqnarray}
where the summation is taken over 4-dimensional vector
$a_\mu = (n_0T, n_1 L, n_2 L, n_3 L)$ with integers $n_i$'s.
Truncation above at $n_{max}=5$ already shows a good convergence
when $M>$ 200 MeV and  $L=T/2=2$ fm, for example.

In this theory, and for the cases we will consider,  one can express $\bar{G}(x,0,0)$ 
in terms of $\bar{\Delta}(x,M^2)$:
%%%%%%%%% Eq Gbar with Delta %%%%%%%%%%%%%%%%%%%%%%%
\begin{eqnarray}
\bar{G}(x,0,0)
&=& \frac{1}{3}\left[
A  \bar{\Delta}\left(x,M^2_{\eta}\right)
+B \bar{\Delta}(x,0) +
C\;\partial_{M^2}\bar{\Delta}(x,0)  \right],
\end{eqnarray}
and, therefore,
%%%%%%%%% Eq r(t) %%%%%%%%%%%%%%%%%%%%%%%%%%%%%%%%%%
\begin{eqnarray}
r(t)&=&\int d^3 x \;\bar{G}(x,0,0)\nonumber\\
&=&\frac{1}{3}\left[
A \left(\frac{\cosh (M_\eta (t-T/2))}{2M_\eta \sinh (M_\eta T/2)}-
\frac{1}{M^2_\eta T}\right)
+BTh_1(t/T)+CT^3h_2(t/T)
\right]\nonumber\\
&=& \frac{1}{3}\left[
BTh_1(t/T)+CT^3h_2(t/T)-\frac{A}{M^2_\eta T}+{\cal O}(e^{-M_\eta t})
\right].
\label{eq:rt}
\end{eqnarray}
%%%%%%%%%%%%%%%%%%%%%
%%%%%%%%%%%%%%%%%%%%% should be moved.
where
%%%%%%%%% h_2 %%%%%%%%%%%%%%%
\begin{eqnarray}
h_2(t/T)\equiv \frac{1}{24}\left[
\left(\frac{t}{T}\right)^2
\left(\frac{t}{T}-1\right)^2-\frac{1}{30}, 
\right].
\end{eqnarray}
and $A, B, C, M_{\eta}$ are functions of the p-regime masses only. The term proportional to $C$ only appears in the case with $N_l=0$.
%partially quenched limit $N_l \rightarrow 0$ and brings in new UV divergences, which are reabsorbed in the condensate, $\Sigma$ (it gets renormalized in the quenched theory, as is well known). 
%After renormalization, we have \cite{Damgaard:2001xr}
With this set up, one obtains
%%%%%%%%% Eq delta %%%%%%%%%%%%%%%%%%%%%%%%%%%%%%%%%
\begin{eqnarray}
\bar{\Delta}(0,0)&=& -\frac{\beta_1}{\sqrt{V}},\;\;\;
\partial_{M^2}\bar{\Delta}(0,M^2)|_{M^2=0}
= -\frac{1}{16\pi^2}\ln \mu_{sub}^2 V^{1/2}
-\beta_2, 
\end{eqnarray}
where $\beta_1$ and $\beta_2$ are the usual shape coefficients
and $\mu_{sub}$ (=0.77GeV in this section) is the subtraction scale.

With this input,one can now calculate meson correlators on the basis
of our expressions.
In the following,  we will give two examples where in both we let 
the volume size be given by $L=2$ fm.
One is the case where the physical up and down quarks are 
in the $\epsilon$-regime, $i.e.$, $N_l=2$ and $N_h=1$.
The other is the case with rather heavy sea quark masses, $i.e.$,
$N_h=3$, but the valence quarks are taken to
the $\epsilon$-regime.

As seen below, %one should note that 
the 1-loop corrections 
to the condensate and decay constant are considerable
even in the limit $V\to \infty$ because of large strange quark mass.
Recently, it has been argued that $N_f=$2+1 flavor ChPT at NLO
may have difficulty in fitting lattice QCD data
\cite{Allton:2008pn, Aoki:2008sm}.
It is clearly important to check whether
NNLO contributions are essential for analyzing such lattice
results at the physical $s$-quark mass,
or if the strange quark is simply out of the region where
ChPT provides a useful expansion. 
If the latter case is true, one would need 
to integrate the strange quark out and
use an "effective" $N_f=2$ ChPT. 
In this paper, we do not wish to address this issue 
and hence just give the NLO formulae for the $N_f=$2+1 theory. Even
in the $N_f=2$ theory one may be interested in keeping the $u$
and $d$ quarks in the $p$-regime, while taking the valence
quarks masses to the $\epsilon$-regime. Our formulas given in
this paper easily extend to that case, but we do not display them here.

\subsection{The case with $N_l=2$, $N_h=1$}

Let us first choose $m_u=m_d=$ 2 MeV, $m_s=$ 110 MeV,
where the physical pions are certainly in the $\epsilon$-regime
in a volume as small as $L=$ 2 fm.

In this case, the coefficients of eq.~(\ref{eq:rt}) are
%%%%%%%% A, B, C %%%%%%%%%%%%%%%%
\begin{eqnarray}
A=-\frac{1}{2},\;\;\; B=\frac{3}{2},\;\;\; C=0,\;\;\;\;\;
%\end{eqnarray}
%%%%%%%% r(t) %%%%%%%%%%%
%\begin{eqnarray}
r(t)=\frac{1}{2}Th_1(t/T)+\frac{1}{6M^2_{\eta}T}, 
\end{eqnarray}
where $M^2_{\eta} = {2 \over 3} M_{ss}^2= {4 m_s \Sigma \over 3 F^2}$. 
The 1-loop corrections to the condensate and decay constant are
then given by
%%%%%%% Sigmatilde Ftilde %%%%%%%
\begin{eqnarray}
\frac{\tilde{\Sigma}}{\Sigma}
&=&1-\frac{1}{F^2}\left[
-\frac{3\beta_1}{2\sqrt{V}}
+\frac{M^2_K}{16\pi^2}\ln \frac{M^2_K}{\mu^2_{sub}}
+\frac{M^2_\eta}{96\pi^2}\ln \frac{M^2_\eta}{\mu^2_{sub}}
-\frac{1}{6M_\eta^2 V}-32L^r_6(\mu_{sub}) M^2_K %+g_1(M^2_K)
\right],\nonumber\\
\frac{\tilde{F}}{F}
&=&1-\frac{1}{2F^2}\left[
-\frac{2\beta_1}{\sqrt{V}}
+\frac{M^2_K}{16\pi^2}\ln \frac{M^2_K}{\mu^2_{sub}}
-16L^r_4(\mu_{sub}) M^2_K %+g_1(M^2_K)
\right],
\end{eqnarray}
where $M^2_K=m_s \Sigma/F^2$, and we have neglected 
exponentially small $g_1(M^2_K)$ and $g_1(M^2_\eta)$ 
($< (1{\rm MeV})^2$). 
One can ignore $k^s_{00}(M^2_K)$, too.
$\mu_{sub}=770 {\rm MeV}$ is what we have taken as the subtraction scale.
%and
%%%%%%%% Gbar(0,0,0) %%%%%%%%%%%
%\begin{eqnarray}
%\bar{G}(0,0,0)&=-\frac{\beta_1}{2\sqrt{V}}
%-\frac{1}{16\pi^2}M^2_{\eta}\ln \frac{M^2_\eta}{\mu^2}+\frac{1}{M^2_\eta V}
%=(\mbox{53 MeV})^2.
%\end{eqnarray}
%The contribution from the Kaon ($M^2_K=m_s \Sigma/F^2$) is
%$\Delta(0,M^2_k)\sim (\mbox{44MeV})^2$, where
%one finds that the finite volume dependent part of $\Delta(0,M^2_K)$
%is numerically very small and 
In the case with $L=T/2=2$fm 
(where $\beta_1=0.0836$), one obtains $\tilde{\Sigma}=1.3\Sigma$ and
$\tilde{F}=1.2F$, respectively.

For the zero-mode integral, we use the partition function
%%%%%%%%%%%%%% Eq (n, m)=(1, N_f+1) partition function deg  %%%%%%%%%%
\begin{eqnarray}
\label{eq:zero-mode-3flavor}
\mathcal{Z}^\nu_{1,1+(N_l=2)}(\mu_b |\mu_v ,\mu)
&=&
\frac{1}{2(\mu^{2}-\mu_v^2)^2}
\nonumber\\
&&\hspace{-1in}
\times \det \left(
\begin{array}{cccc}
K_\nu(\mu_{b}) & I_\nu(\mu_v) & I_\nu(\mu) & 
\mu^{-1}I_{\nu-1}(\mu) \\
-\mu_{b} K_{\nu+1}(\mu_{b}) & \mu_v I_{\nu+1}(\mu_v) & \mu I_{\nu+1}(\mu) 
& I_{\nu}(\mu)\\
\mu_{b}^2 K_{\nu+2}(\mu_{b}) & \mu_v^2I_{\nu+2}(\mu_v) & 
\mu^2I_{\nu+2}(\mu) 
& \mu I_{\nu+1}(\mu) \\
-\mu_{b}^3 K_{\nu+3}(\mu_{b}) & \mu_v^3I_{\nu+3}(\mu_v) & 
\mu^3I_{\nu+3}(\mu) 
& \mu^2 I_{\nu+2}(\mu) \\
\end{array}
\right),
\end{eqnarray}
where $\mu=m_u\Sigma V=m_d \Sigma V$.

When the valence masses are degenerate, we use
%%%%% Eq K,J Nl=2, degenerate %%%%%%%%%%%%%%%%
\begin{eqnarray}
\mathcal{K}^0_{+}&=&
2\frac{\partial_{\mu_v} \Sigma^{\rm PQ}_\nu(\mu_v, \mu)}{\Sigma},
\;\;\;
\mathcal{K}^0_{-}=-2\frac{\Sigma^{\rm PQ}_\nu(\mu_v, \mu)}{\mu_v\Sigma},
\nonumber\\
\mathcal{K}^1_{\pm}&=&1\pm \left(D^{\rm PQ}_\nu (\mu_v,\mu_v,\mu)
+\frac{\nu^2}{\mu_v^2}\right),\;\;\;
\mathcal{J}^0_{\pm}=1\pm \left(D^{\rm PQ}_\nu (\mu_v,\mu_v,\mu)
-\frac{\nu^2}{\mu_v^2}\right),
\nonumber\\
\mathcal{J}^1_{+}&=&8m_v  \frac{\Sigma^{\rm PQ}_\nu(\mu_v, \mu)}{\Sigma}
- \frac{4}{\Sigma V}(\mathcal{J}^0_+-\mathcal{J}^0_-).
\end{eqnarray}

We now present the explicit form of the 
correlators for the pseudoscalar and axial vector channels
with $m_v=m_{v^\prime}$,
%%%%%%%% Eq correlators example Nl=2, Nh=1 %%%%%%%%%%%%
\begin{eqnarray}
\mathcal{P}^c_{vv^\prime}(t)&=&
L^3\left(\frac{\tilde{\Sigma}^2}{2\tilde{\mu}_v}
\frac{\Sigma^{\rm PQ}_\nu(\tilde{\mu_v},\tilde{\mu})}{\Sigma}
-\frac{\Sigma^2}{6F^2M^2_\eta V}
\frac{\partial_{\mu_v}\Sigma^{\rm PQ}_\nu(\mu_v,\mu)}{\Sigma}\right)\nonumber\\
&&+\frac{\Sigma^2}{2F^2}
\left(1+D^{\rm PQ}_\nu(\mu_v,\mu_v,\mu)+\frac{\nu^2}{\mu_v^2}
-\frac{\partial_{\mu_v}\Sigma^{\rm PQ}_\nu(\mu_v,\mu)}{\Sigma}\right)Th_1(t/T),\\
\mathcal{A}^c_{vv^\prime}(t)&=&
-\frac{\tilde{F}^2}{2T}
\left(1+D^{\rm PQ}_\nu(\tilde{\mu_v},\tilde{\mu_v},\tilde{\mu})
-\frac{\nu^2}{\tilde{\mu_v}^2}\right)
+\frac{Tk_{00}}{V}
\left(1-D^{\rm PQ}_\nu(\mu_v,\mu_v,\mu)+\frac{\nu^2}{\mu_v^2}\right)
\nonumber\\
&&-\frac{2\mu_v}{V}
\frac{\Sigma^{\rm PQ}_\nu(\mu_v,\mu)}{\Sigma}
Th_1(t/T),
\end{eqnarray}
where $\tilde{\mu}_i = m_i \tilde{\Sigma}V$.
We plot these correlators in Figs. \ref{fig:corrPPNl2}
and \ref{fig:corrAANl2} using $k_{00}=0.08331$ for this case.

\subsection{$N_l=0$, $N_h=3$}
As the second example, let us consider the case with
$m_u=m_d=$ 30 MeV, $m_s=$ 110 MeV while the valence
quark masses are taken to be very light, $m_v={\cal O}(1)$ MeV.
In this case, all the sea quarks are in the $p$-regime
and we therefore have $N_l=0$ and $N_h=3$. In this case, we have
%%%%% Eq 2+1 alpha beta gamma %%%%%%%%%%%%%%%%%%
\begin{eqnarray}
A &=& -\frac{2(M^2_{ud}-M^2_{ss})^2}
{(M^2_{ud}+2M^2_{ss})^2},\;\;\;
B = 1 + \frac{2(M_{ud}^2-M_{ss}^2)^2}{(M^2_{ud}+2M^2_{ss})^2},\;\;\;
%\nonumber\\
C=- 3 \frac{M^2_{ud}M^2_{ss}}{M^2_{ud}+2M^2_{ss}}, 
\end{eqnarray}
while $M_\eta^2 = (M^2_{ud}+2M^2_{ss})/3$, where
$M^2_{ud}=(m_u+m_d)\Sigma/F^2$, $M^2_{ss}=2m_s\Sigma /F^2$. 
Note that a double pole contribution now appears in $r(t)$
as a partial quenching artifact since $C \neq 0$. 
%We take the renormalization scale of the condensate to be the same as that of the higher order couplings $L_4^r$ and $L_6^r$. 
%In the following formulae therefore $\Sigma$ represents $\Sigma^r(\mu_{sub})$. 

The 1-loop corrections to the condensate and decay constant
in this case are
%%%%%%%%% Eq Sigmatilde Ftilde N_h=3 %%%%%%%%%%%%%%%
\begin{eqnarray}
\frac{\tilde{\Sigma}}{\Sigma}
&=&1-\frac{1}{F^2}\left[
\frac{2M^2_\pi}{16\pi^2}\ln \frac{M^2_\pi}{\mu^2_{sub}}
+2g_1(M^2_\pi)
+\frac{M^2_K}{16\pi^2}\ln \frac{M^2_K}{\mu^2_{sub}}\right.
\nonumber\\
&&-\frac{A}{3}\left(
\frac{M^2_\eta}{16\pi^2}\ln \frac{M^2_\eta}{\mu^2_{sub}}
-\frac{1}{M_\eta^2 V}\right)
+B\frac{\beta_1}{3\sqrt{V}}+\frac{C}{3}
\left(\frac{\ln \mu^2_{sub} V^{1/2}}{16\pi^2}+\beta_2\right)
\nonumber\\&&
\left.-32L^r_6(\mu_{sub}) (M^2_\pi+M^2_K) %+g_1(M^2_K)
\right],\nonumber\\
\frac{\tilde{F}}{F}
&=&1-\frac{1}{2F^2}\left[
\frac{2M^2_\pi}{16\pi^2}\ln \frac{M^2_\pi}{\mu^2_{sub}}
+2g_1(M^2_\pi)
+\frac{M^2_K}{16\pi^2}\ln \frac{M^2_K}{\mu^2_{sub}}
-16L^r_4(\mu_{sub}) (M^2_{\pi}+M^2_K) %+g_1(M^2_K)
\right],\nonumber\\
\end{eqnarray}
where $M_\pi^2=(m_u+m_d)\Sigma/F^2$ denotes the pion mass.
Again we set $\mu_{sub}=770$ MeV.
In this case, the corrections are uncomfortably large:
$\tilde{\Sigma}=1.5\Sigma$ and $\tilde{F}=1.2F$.

The zero-mode partition function for $N_l=0$ is given by
%%%%%%%%%%%%%% Eq (n, m)=(1, N_f+1) partition function deg  %%%%%%%%%%
\begin{eqnarray}
\label{eq:zero-mode-Nl=0}
\mathcal{Z}^\nu_{1,1+(N_l=0)}(\mu_b |\mu_v )
&=&
%\frac{1}{2(\mu^{2}-\mu_v^2)^2}
%\nonumber\\
%&&\hspace{-1in}
%\times 
\det \left(
\begin{array}{cc}
K_\nu(\mu_{b}) & I_\nu(\mu_v) \\%& I_\nu(\mu) & \mu^{-1}I_{\nu-1}(\mu) \\
-\mu_{b} K_{\nu+1}(\mu_{b}) & \mu_v I_{\nu+1}(\mu_v) \\%& \mu I_{\nu+1}(\mu) & I_{\nu}(\mu)\\
%\mu_{b}^2 K_{\nu+2}(\mu_{b}) & \mu_v^2I_{\nu+2}(\mu_v) \\%& \mu^2I_{\nu+2}(\mu) & \mu I_{\nu+1}(\mu) \\
%-\mu_{b}^3 K_{\nu+3}(\mu_{b}) & \mu_v^3I_{\nu+3}(\mu_v) \\%& \mu^3I_{\nu+3}(\mu) & \mu^2 I_{\nu+2}(\mu) \\
\end{array}
\right).
\end{eqnarray}

When the valence masses are degenerate, we use
%%%%% Eq K,J Nl=2, degenerate %%%%%%%%%%%%%%%%
\begin{eqnarray}
\mathcal{K}^0_{+}&=&
2\frac{\partial_{\mu_v} \Sigma^{\rm FQ}_\nu(\mu_v)}{\Sigma},
\;\;\;
\mathcal{K}^0_{-}=-2\frac{\Sigma^{\rm FQ}_\nu(\mu_v)}{\mu_v\Sigma},\;\;\;
\mathcal{K}^1_{+}=2+2\frac{\nu^2}{\mu_v^2},\nonumber\\
\mathcal{J}^0_+&=&2,\;\;\;\mathcal{J}^0_-=0,\;\;\;\;
\mathcal{J}^1_{+}=8m_v  \frac{\Sigma^{\rm FQ}_\nu(\mu_v)}{\Sigma}.
%- \frac{4}{\Sigma V}(\mathcal{J}^0_+-\mathcal{J}^0_-).
\end{eqnarray}

Here we present the  correlators for the pseudoscalar and axial vector
channels ($m_v=m_{v^\prime}$), 
%%%%%%%% Eq correlators example Nl=2, Nh=1 %%%%%%%%%%%%
\begin{eqnarray}
\mathcal{P}^c_{vv^\prime}(t)&=&
L^3\left(\frac{\tilde{\Sigma}^2}{2\tilde{\mu}_v}
\frac{\Sigma^{\rm FQ}_\nu(\tilde{\mu_v})}{\Sigma}
+\frac{A\Sigma^2}{3F^2M^2_\eta V}
\frac{\partial_{\mu_v}\Sigma^{\rm FQ}_\nu(\mu_v)}{\Sigma}\right)\nonumber\\
&&+\frac{\Sigma^2}{2F^2}
\left(2+2\frac{\nu^2}{\mu_v^2}
%1+D^{\rm PQ}_\nu(\mu_v,\mu_v,\mu)+\frac{\nu^2}{\mu_v^2}
-\frac{2B}{3}\frac{\partial_{\mu_v}\Sigma^{\rm FQ}_\nu(\mu_v)}{\Sigma}\right)Th_1(t/T)
\nonumber\\
&&-\frac{\Sigma^2}{2F^2}
\left(\frac{2C}{3}
\frac{\partial_{\mu_v}\Sigma^{\rm FQ}_\nu(\mu_v)}{\Sigma}\right)T^3h_2(t/T)
,\\
\mathcal{A}^c_{vv^\prime}(t)&=&
-\frac{\tilde{F}^2}{T}
%+\frac{Tk_{00}}{V}
%\left(1-D^{\rm PQ}_\nu(\mu_v,\mu_v,\mu)+\frac{\nu^2}{\mu_v^2}\right)
%\nonumber\\
%&&
-\frac{2\mu_v}{V}
\frac{\Sigma^{\rm FQ}_\nu(\mu_v)}{\Sigma}Th_1(t/T),
\end{eqnarray}
where $\tilde{\mu}_i = m_i \tilde{\Sigma}V$.
We plot these correlators in Figs. \ref{fig:corrPPNl0}
 and \ref{fig:corrAANl0}.

\FIGURE[p]{
%\begin{figure}[p]\centering
%\includegraphics[width=13cm]{corrPPNl2mv.eps}
%\includegraphics[width=13cm]{corrPPNl2nu.eps}
\epsfig{file=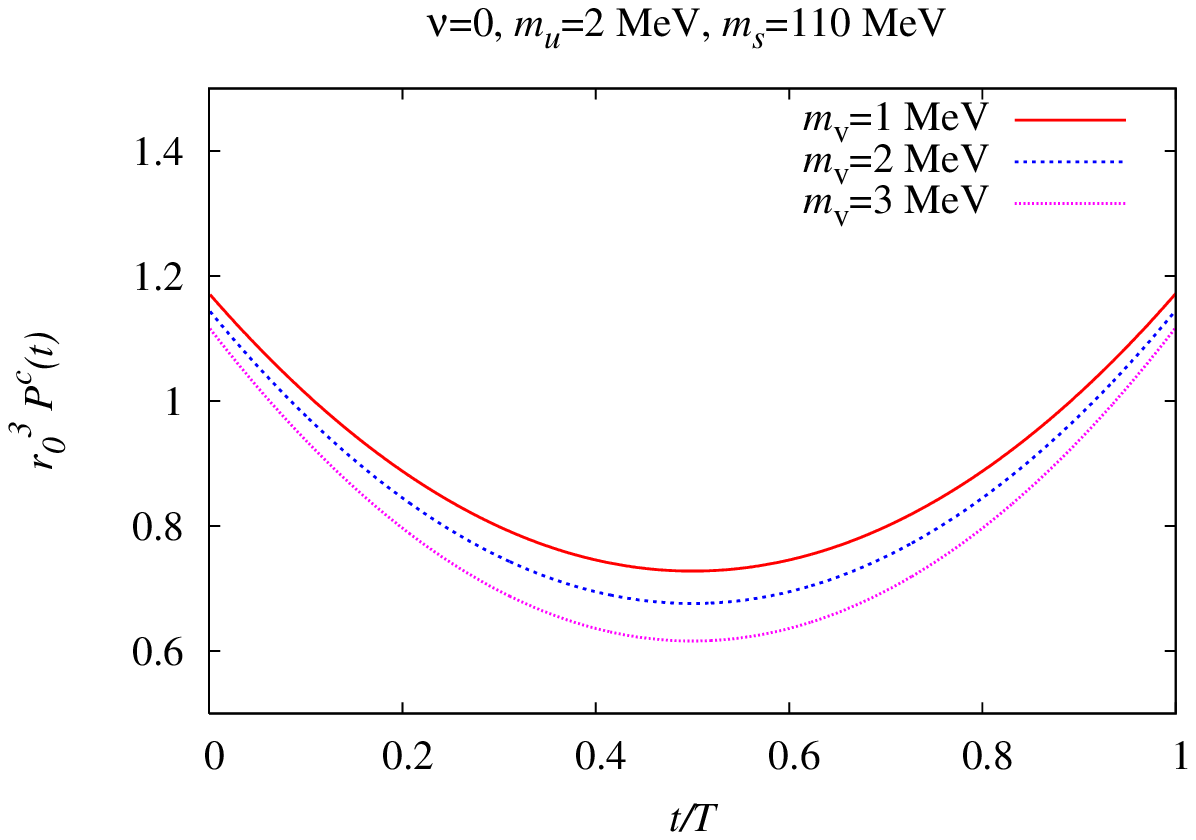, width=13cm}
\epsfig{file=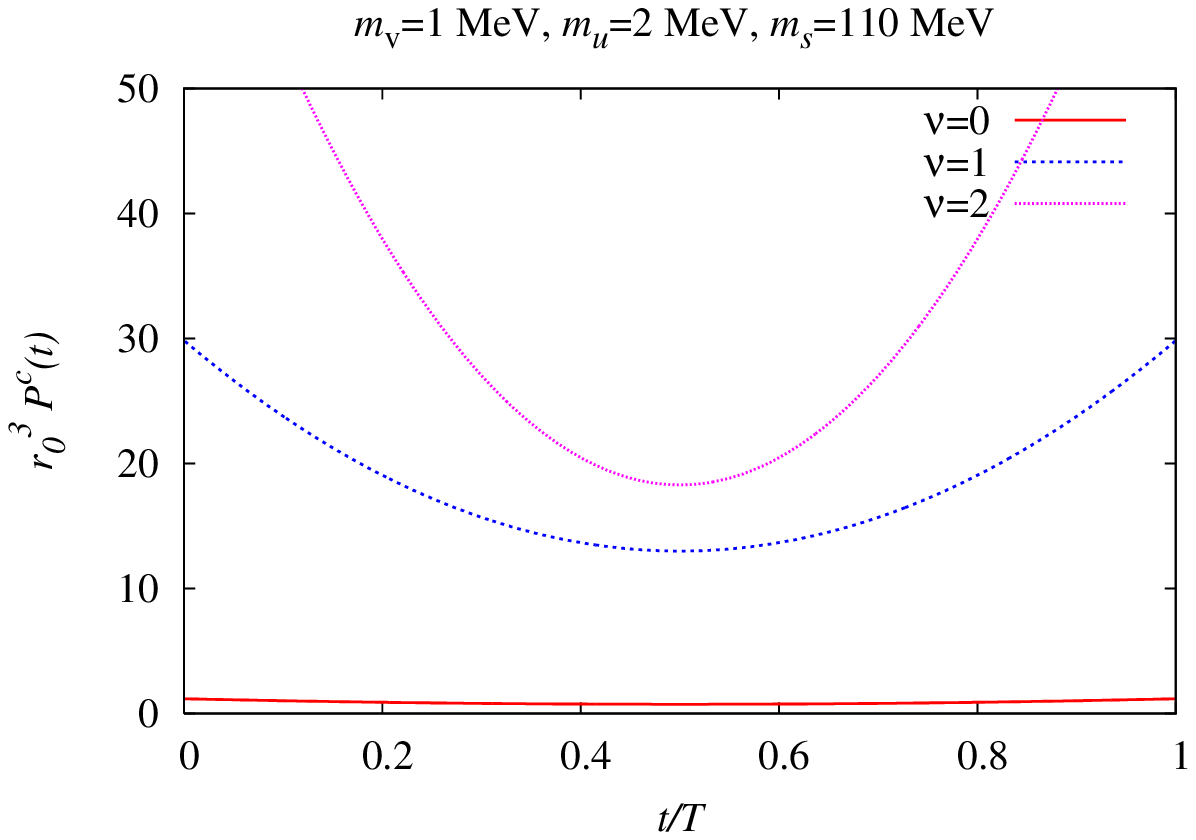, width=13cm}
\caption{\label{fig:corrPPNl2}
The pseudoscalar correlators with $m_v=1$-3 MeV,
$m_u=m_d=2$ MeV and $m_s=110$ MeV
in a sector of trivial topology, $\nu=0$ (top)
and  in sectors of $\nu=0$-2 for fixed $m_v=1$ MeV (bottom).
We set $L=T/2=2$ fm and the correlators are 
normalized by the Sommer scale $r_0=0.49$ fm \cite{Necco:2001xg}.
}
}
%\end{figure}

\FIGURE[p]{
%\begin{figure}[p]\centering
%\includegraphics[width=13cm]{corrAANl2mv.eps}
%\includegraphics[width=13cm]{corrAANl2nu.eps}
\epsfig{file=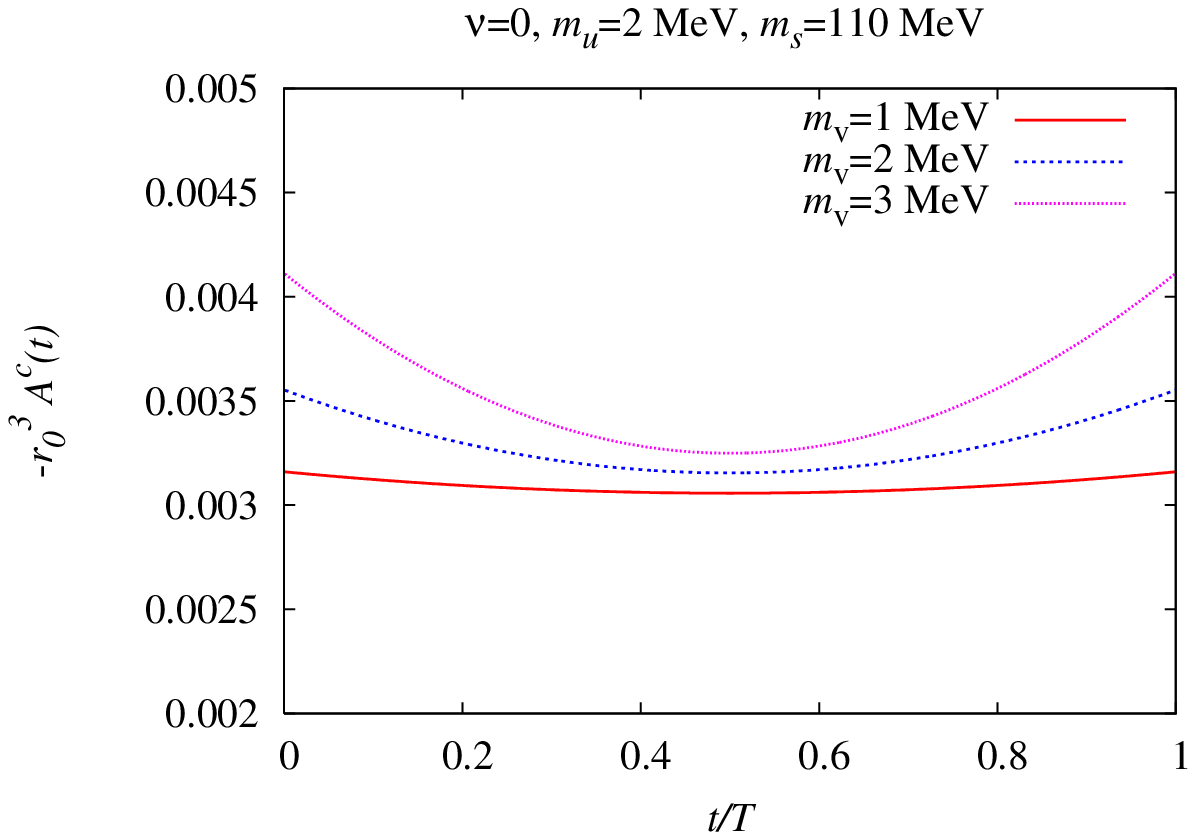, width=13cm}
\epsfig{file=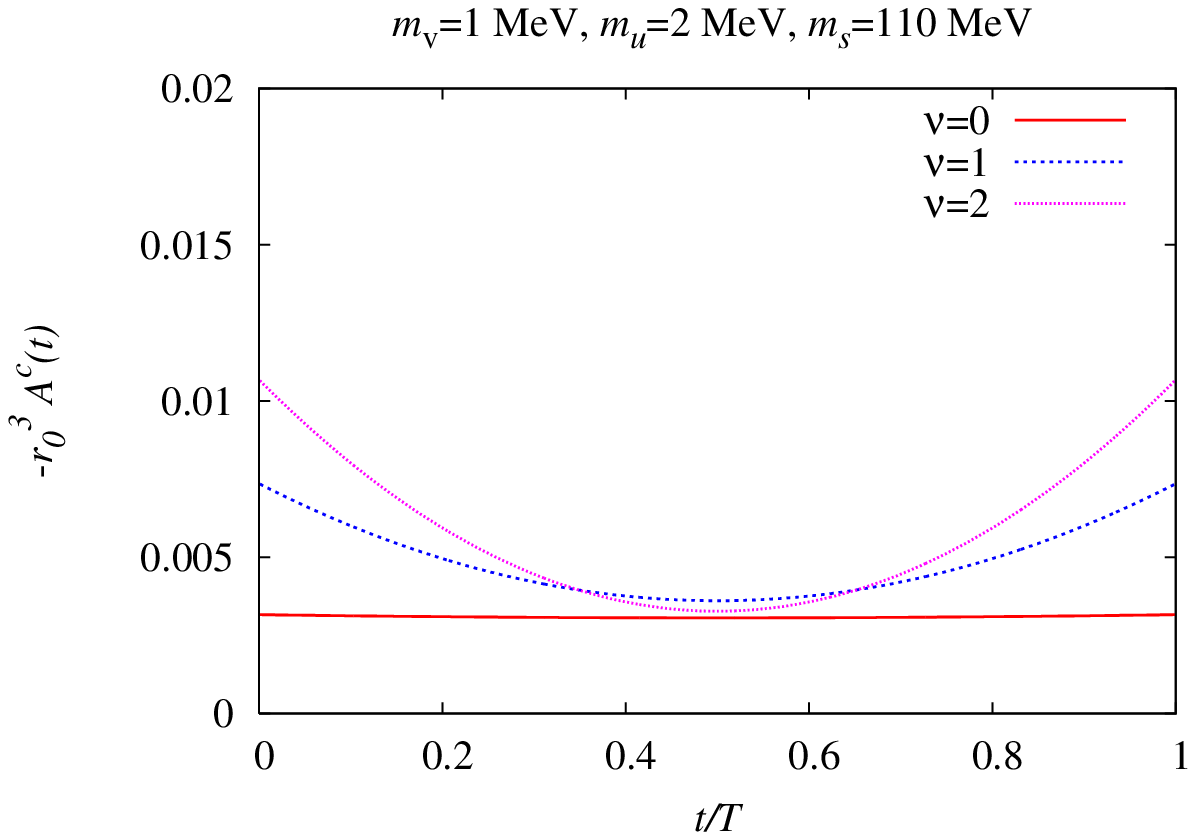, width=13cm}
\caption{\label{fig:corrAANl2}
The axial correlators for same parameter values
as in Fig. \ref{fig:corrPPNl2}.
}
%\end{figure}
}

\FIGURE[p]{
%\begin{figure*}[p]
%  \centering
%  \includegraphics[width=13cm]{corrPPNl0mv.eps}
%  \includegraphics[width=13cm]{corrPPNl0nu.eps}
\epsfig{file=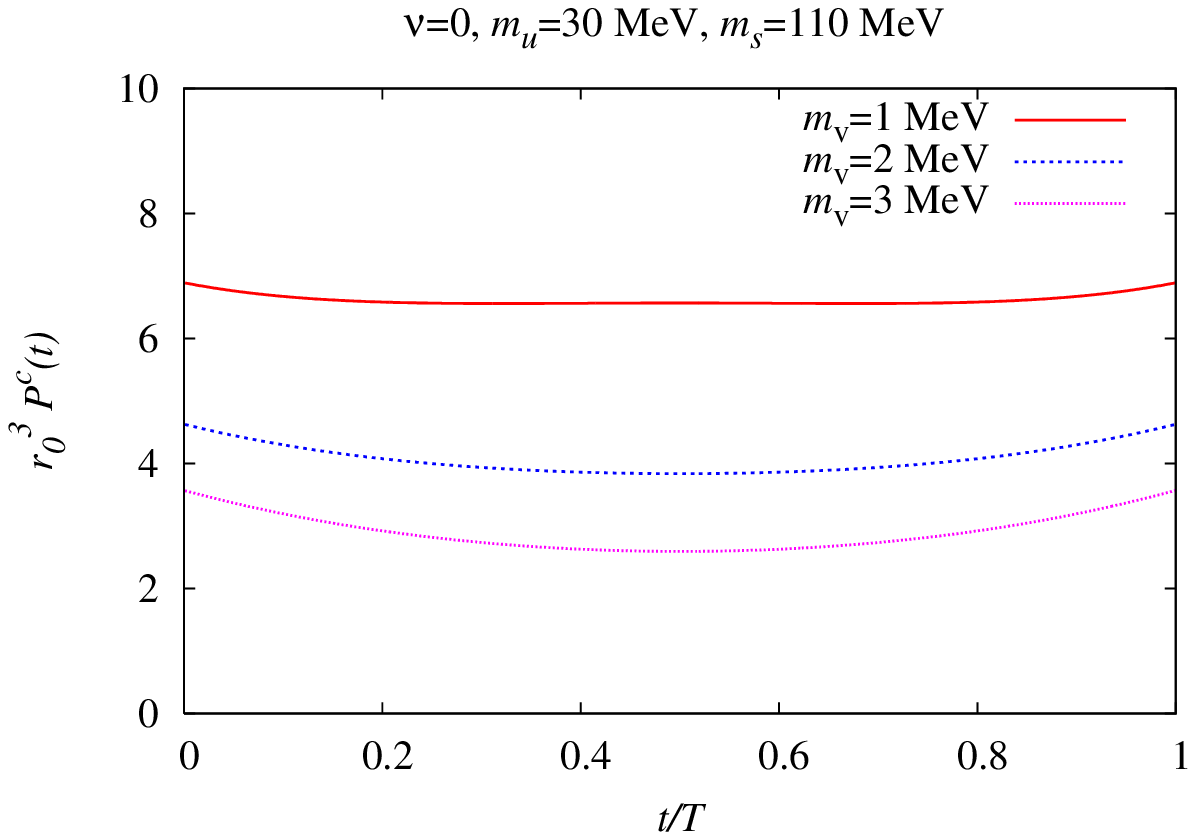, width=13cm}
\epsfig{file=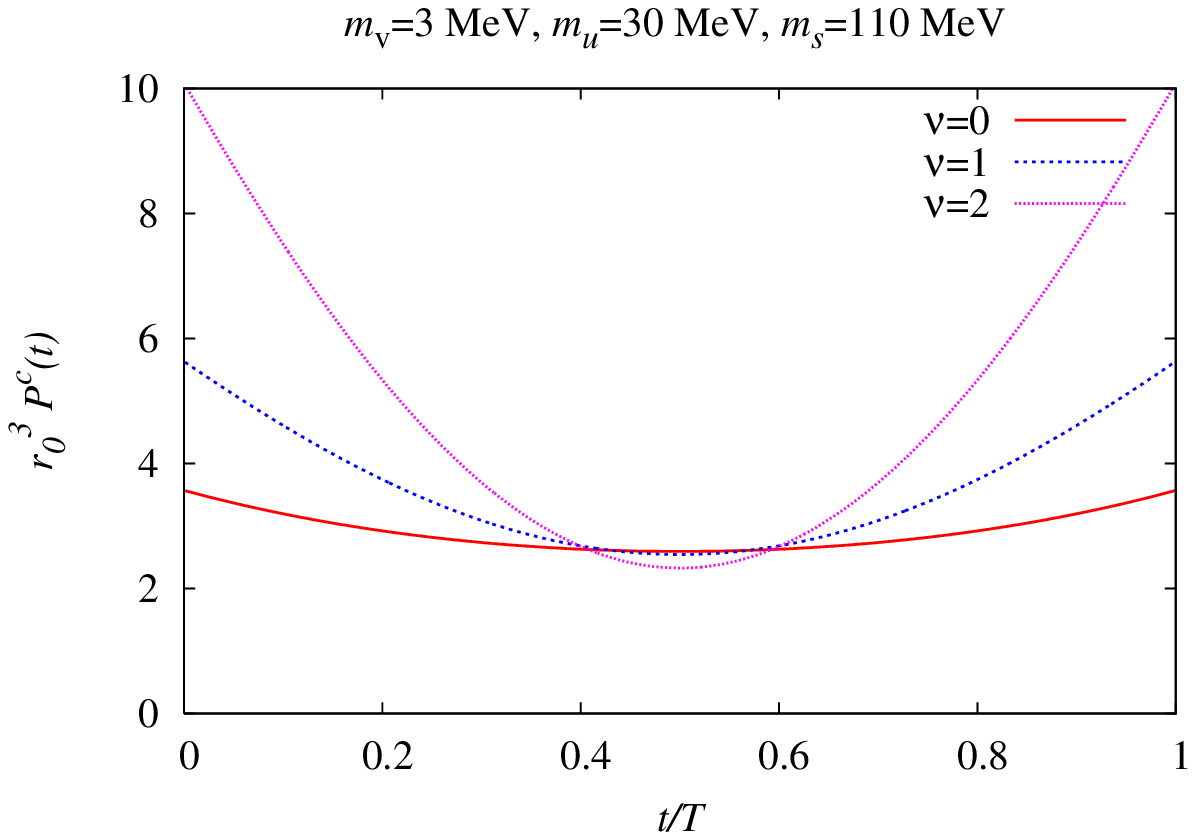, width=13cm}
  \caption{
    The pseudoscalar correlators with $m_v=1$-3 MeV,
    $m_u=m_d=30$ MeV and $m_s=110$ MeV
    in a sector of trivial topology, $\nu=0$ (top)
    and in sectors of $\nu=0$-2 for fixed $m_v=3$ MeV (bottom).
%    We set $L=T/2=2$ fm and the correlators are 
%    normalized by the Sommer scale $r_0=0.49$ fm \cite{Necco:2001xg}.
  \label{fig:corrPPNl0}}
%\end{figure*}
}

\FIGURE[p]{
%\begin{figure}[p]
%  \centering
%  \includegraphics[width=13cm]{corrAANl0mv.eps}
%  \includegraphics[width=13cm]{corrAANl0nu.eps}
\epsfig{file=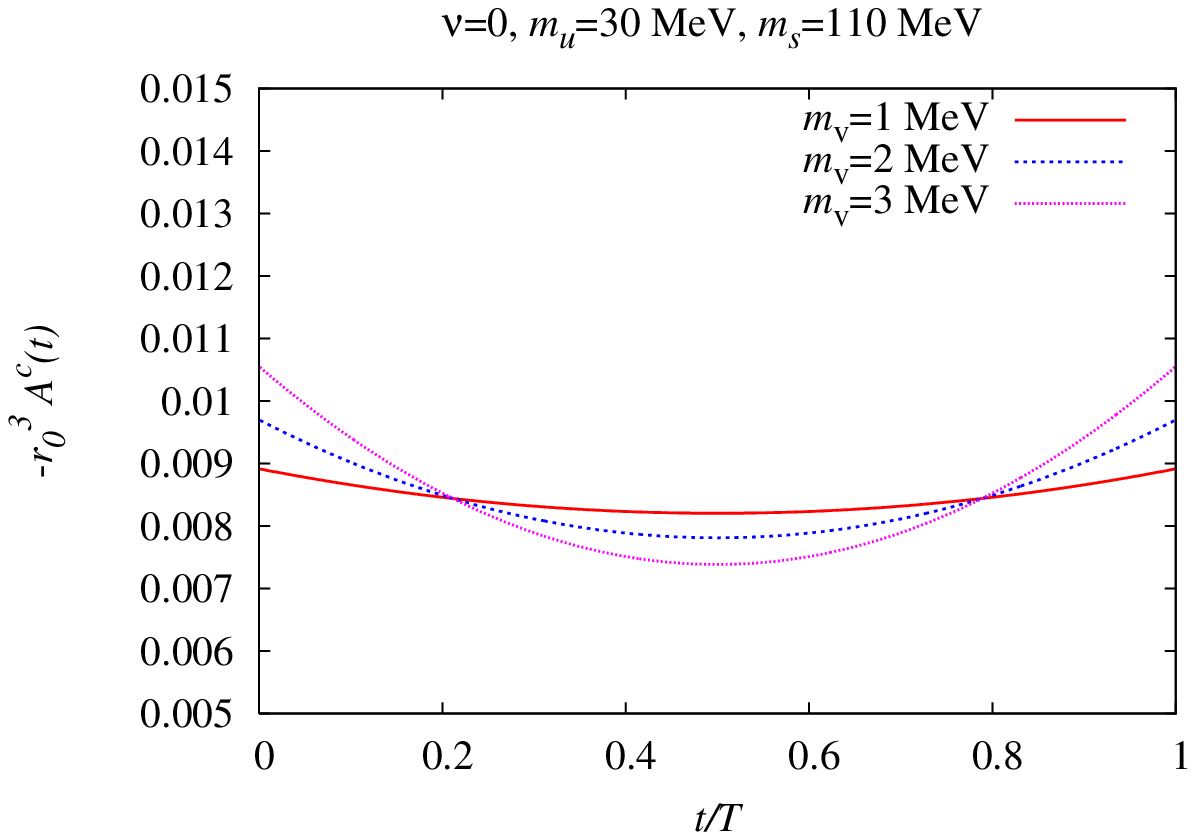, width=13cm}
\epsfig{file=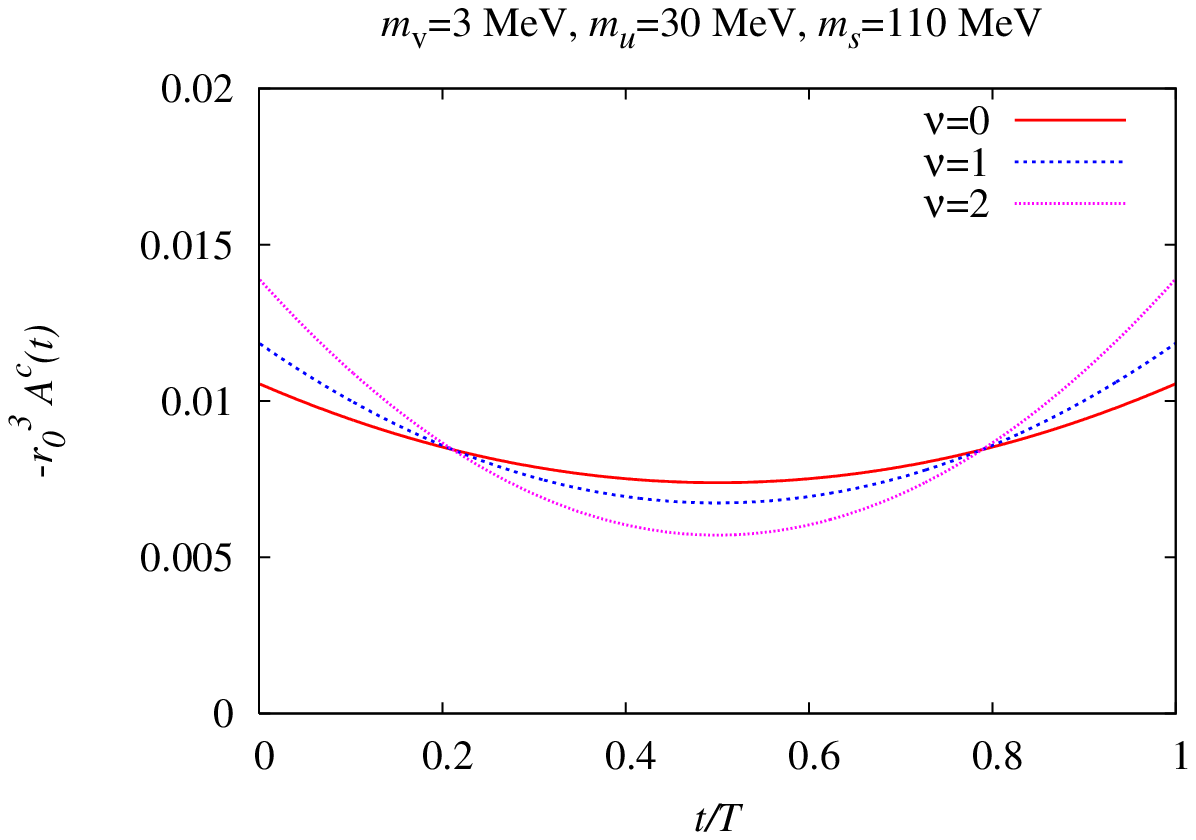, width=13cm}
  \caption{
    The axial correlators for the same parameter values
    as in Fig. \ref{fig:corrPPNl0}. 
  \label{fig:corrAANl0}}
%\end{figure}
}

%% file: sec6.tex
%%%%%%%%%%%%%%%%%%%%%%%%%%%%%%%%%%%%%%%%%%%%%%%%%%%
%%%%%%%%%%%%%%%%%%%%%%%%%%%%%%%%%%%%%%%%%%%%%%%%%%%
%%%%%%%%%%%%%%%%%%%%%%%%%%%%%%%%%%%%%%%%%%%%%%%%%%%
\section{Conclusions}
\label{sec:conclusions}
\setcounter{equation}{0}
%%%%%%%%%%%%%%%%%%%%%%%%%%%%%%%%%%%%%%%%%%%%%%%%%%%
%%%%%%%%%%%%%%%%%%%%%%%%%%%%%%%%%%%%%%%%%%%%%%%%%%%
%%%%%%%%%%%%%%%%%%%%%%%%%%%%%%%%%%%%%%%%%%%%%%%%%%%

We have developed a new scheme of calculations
for chiral perturbation theory 
with non-degenerate quark masses 
in finite volume.
With our new counting rules separating 
$N_l$ light quarks in the $\epsilon$-regime
and the other $N_h$ quarks in the $p$-regime, 
we have calculated the meson correlators in
various channels: pseudoscalar, scalar,
vector and axial vector.
We have also calculated the disconnected contributions
for the pseudoscalar and scalar channels.

With the help of the replica method,
we have also extended our study to 
the partially quenched case.
Our results are shown to be consistent with all earlier
work in the literature, both in the quenched and full QCD limit,
with degenerate valence quarks.

Our results can be compared to 
lattice QCD simulation with 2+1 flavors
where the up and down quark masses are very
light, but the volume is such that the theory is
in the $\epsilon$-regime with respect to the corresponding
pseudo Nambu-Goldstone bosons.
The two-point functions are useful to
determine the leading low-energy constants,
the chiral condensate $\Sigma$, and the
pion decay constant $F$ in the chiral limit. As we have
demonstrated, the formulae may also be used to extract the
numerical values of higher-order low energy constants
$L_4$ and $L_6$.
This  work can be extended to other observables in the case 
where one valence quark is heavy, $i.e.$ the 
chiral dynamics of kaons. With the new partially quenched 
chiral perturbation theory in this mixed regime one has an
excellent analytical tool with which to explore future
lattice simulations with nearly massless $u$ and $d$ quarks.
It would be most interesting to 
investigate by analytic means also
the region between the two regimes, where $m_\pi L\sim 1$.

%%%%%%%%%%%%%%%%%%%%%%%%%%%%%%%%%
\section*{Acknowledgments}
%%%%%%%%%%%%%%%%%%%%%%%%%%%%%%%%%

The authors thank Silvia Necco for useful information and
comments on $2+1$ flavor ChPT.
PHD and HF would like to thank the members of IFIC for warm hospitality 
during their stay in Valencia. 
FB acknowledges finantial support from the FPU grant AP2005-5201.
The work of PHD was partly supported by the EU network
ENRAGE MRTN-CT-2004-005616.
The work of HF was supported by Nishina Memorial Foundation
and Japan Society for the Promotion of Science.  FB and PH  acknowledge partial financial 
support from the research grants FPA-2007-01678, 
FLAVIAnet and Consolider-Ingenio 2010 Programme CPAN (CSD2007-00042). PH thanks the CERN Theory Division for hospitality while this work was completed. 
%%%%%%%%%%%%%%%%%%% Fabio, Poul, Pilar, will write.

%% file: appendix.tex
\appendix

%%%%%%%%%%%%%%%%%%%%%%%%%%%%%%%%%%%%%%%%%%%%%%%%%%%
%%%%%%%%%%%%%%%%%%%%%%%%%%%%%%%%%%%%%%%%%%%%%%%%%%%
%%%%%%%%%%%%%%%%%%%%%%%%%%%%%%%%%%%%%%%%%%%%%%%%%%%
\section{
%Zero-momentum projection of $\bar{\Delta}$'s
Space-integrals involving propagators} 
\label{app:zero-momentum}
\setcounter{equation}{0}
%%%%%%%%%%%%%%%%%%%%%%%%%%%%%%%%%%%%%%%%%%%%%%%%%%%
%%%%%%%%%%%%%%%%%%%%%%%%%%%%%%%%%%%%%%%%%%%%%%%%%%%
%%%%%%%%%%%%%%%%%%%%%%%%%%%%%%%%%%%%%%%%%%%%%%%%%%%

In this appendix, we list several useful formulae for
zero-momentum projection (or, equivalently, 
3-dimensional space integrals)
of functions expressed by $\bar{\Delta}(x,M^2)$.
%at finite volume $V=L^3T$:
%%%%%%%%%%%%%%%%%%%%%%%%%%%%%%%%%%%%%%
%%%%%%%%%% Eq delta bar
%\begin{eqnarray}
%\label{eq:deltabar}
%\bar{\Delta}(x;M^2_{ij})=
%\frac{1}{V}\sum_{p\neq 0}\frac{e^{ipx}}{p^2+M^2_{ij}},
%\end{eqnarray}
%where the summation is taken over the 4-momentum 
%%%%%%%%%%%%%%%%%%%%%%%%%%%%%%%%%%%%%%
%%%%%%%%%% Eq 4-momenta
%\begin{equation}
%p=(2\pi n_t/T, 2\pi n_x/L, 2\pi n_y/L, 2\pi n_z/L),
%\end{equation}
%with integers $n_i$'s.

A useful identity is
%%%%%%%%%%%%%%%%%%%%%%%%%%%%%%%%%%%%%%
%%%%%%%%%% Eq summation formula
\begin{eqnarray}
\hspace{-0.25in}
\sum_{n}\frac{g\left(\frac{2\pi n}{L}\right)
e^{i\frac{2\pi n}{L}x}}{\left(\frac{2\pi n}{L}\right)^2+M^2}
=\frac{L}{4M}\frac{1}{\sinh\left(\frac{ML}{2}\right)}
\left[
g(iM)e^{-M(x-L/2)}+g(-iM)e^{M(x-L/2)}
\right],\nonumber\\
\end{eqnarray}
which holds for an arbitrary regular function $g(p)$.
The zero-mode projection of $\bar{\Delta}(x,M^2)$,
for example,
can be easily derived by setting $g=1$;
%%%%%%%%%%%%%%%%%%%%%%%%%%%%%%%%%%%%%%
%%%%%%%%%% Eq deltabar integration
\begin{eqnarray}
\int d^3 x \bar{\Delta}(x,M^2)
%&=&\frac{L^3}{V}\sum_{n_t\neq 0}\frac{e^{i\frac{2\pi n_t}{T}t}}
%{\left(\frac{2\pi n_t}{T}\right)^2+M^2_{ij}}\nonumber\\
&=&\frac{1}{2M}\frac{\cosh(M(t-T/2))}{\sinh\left(MT/2\right)}
-\frac{1}{M^2T}.
\end{eqnarray}
We are particularly interested in the massless limit:
%%%%%%%%%%%%%%%%%%%%%%%%%%%%%%%%%%%%%%
%%%%%%%%%% Eq deltabar integration massless
\begin{eqnarray}
\int d^3 x \bar{\Delta}(x,0)
&=&
\frac{T}{2}\left(\frac{t}{T}-\frac{1}{2}\right)^2-\frac{T}{24}
\equiv Th_1(t/T).
\end{eqnarray}
It follows its second time derivative is given by
%%%%%%%%%%%%%%%%%%%%%%%%%%%%%%%%%%%%%%
%%%%%%%%%% Eq partial_0^2 deltabar
\begin{eqnarray}
\int d^3 x \partial_0^2\bar{\Delta}(x,M^2)
&=&\frac{M}{2}\frac{\cosh(M(t-T/2))}{\sinh\left(MT/2\right)},\\
\int d^3 x \partial_0^2\bar{\Delta}(x,0)
&=&\frac{1}{T}.
\end{eqnarray}
In this paper, we have also needed the following integral 
involving two $\bar{\Delta}$'s
%%%%%%%%%%%%%%%%%%%%%%%%%%%%%%%%%%%%%%
%%%%%%%%%% Eq two delta in mass term
\begin{eqnarray}
\hspace{-0.2in}
\int d^3 x \int d^4z \partial_0\bar{\Delta}(z-x, 0)
\partial_0 \bar{\Delta}(z, 0)
%&=&
%\int d^3x \int d^4 z \sum_{p,p^\prime\neq 0}
%\frac{-p_0p^\prime_0e^{i(p+p^\prime)z}e^{-ipx}}{V^2p^2p^{\prime 2}}
%\nonumber\\
%&=& 
%\int d^3 x \frac{1}{V}\sum_{p\neq 0}\frac{p_0^2e^{-ipx}}{(p^2)^2}
%\nonumber\\
&=& %\frac{L^3}{V}\sum_{p_0\neq 0}\frac{p_0^2}{p_0^4}e^{-ip_0 x}=
Th_1(t/T),
\end{eqnarray}
and the more non-trivial integral
%%%%%%%%%%%%%%%%%%%%%%%%%%%%%%%%%%%%%%
%%%%%%%%%% Eq two delta most complicated one
\begin{eqnarray}
\label{eq:deltaformula}
\int d^3 x
\left(\partial_0\bar{\Delta}(x,M^2)\partial_0
\bar{\Delta}(x,M^2)
-\bar{\Delta}(x,M^2)\partial^2_0
\bar{\Delta}(x,M^2)\right)\hspace{.7in}
\nonumber\\
=\frac{T}{V}\left(k^s_{00}(M^2)+\frac{1}{M^2T^2}\right)
+\frac{1}{V}\left(
\frac{\cosh(M(t-T/2))}{2M\sinh(MT/2)}
-\frac{1}{M^2T}\right),
\end{eqnarray}
where 
%%%%%%%%%% Eq k^s_00 %%%%%%%%%%%%%%%%%%%%%%%%%%%%
\begin{eqnarray}
k^s_{00}(M^2)\equiv \sum_{\vec{q}=(p_1,p_2,p_3)}
\frac{-1}{4\sinh^2(\sqrt{|\vec{q}|^2+M^2}T/2)}.
\end{eqnarray}
%which needs an appropriate regularization.
%%%%%%%%%%% REF !!!!!!!!!!!!!!!!!!!!!!!!!
The chiral limit of Eq.(\ref{eq:deltaformula})
is given by
\begin{eqnarray}
\label{eq:deltaformula2}
\int d^3 x
\left(\partial_0\bar{\Delta}(x,0)\partial_0
\bar{\Delta}(x,0)
-\bar{\Delta}(x,0)\partial^2_0
\bar{\Delta}(x,0)\right)
\nonumber\\
=\frac{T}{V}k_{00}+\frac{T}{V}h_1(t/T),
\end{eqnarray}
where
%%%%%%%%%% Eq k_00 %%%%%%%%%%%%%%%%%%%%%%%%%%%%%
\begin{equation}
k_{00}\equiv \sum_{\vec{q}=(p_1,p_2,p_3)
\neq 0}\frac{-1}{4\sinh^2(|\vec{q}|T/2)}+
\frac{1}{12},
\end{equation}
becomes now a constant depending only on the shape of the box
\cite{Hasenfratz:1989pk}.

\section{Summary of zero-mode group integrals}
\label{app:Uint}
\setcounter{equation}{0}
%%%%%%%%%%%%%%%%%%%%%%%%%%%%%%%%%%%%%%%%%%%%%%%%%%%
%%%%%%%%%%%%%%%%%%%%%%%%%%%%%%%%%%%%%%%%%%%%%%%%%%%
%%%%%%%%%%%%%%%%%%%%%%%%%%%%%%%%%%%%%%%%%%%%%%%%%%%

Here we summarize the most essential zero-mode group integrals 
which are needed in the general partially quenched case,
see also ref. \cite{Damgaard:2007ep} %Sec.\ref{sec:Uint} 
for additional details.

The zero-mode contribution to the partition function
with $n$ bosons and $m$ fermions is known as seen in 
Eq.(\ref{eq:zero-mode}).
%in closed analytical form for an arbitrary mass matrix
%\cite{Splittorff:2002eb},
%%%%%%%%%%%%% Eq zero-mode partition function %%%%%%%%%%%%%
%\begin{equation}
%%\label{eq:zero-mode}
%\mathcal{Z}^\nu_{n,m}(\{\mu_i\})
%=
%\frac{\det[\mu_i^{j-1}\mathcal{J}_{\nu +j-1}(\mu_i)]_{i,j=1,\cdots n+m}}
%{\prod_{j>i=1}^n(\mu_j^2-\mu_i^2)\prod_{j>i=n+1}^{n+m}(\mu_j^2-\mu_i^2)},
%\end{equation}
%where $\mu_i=m_i\Sigma V$.
%Here $\mathcal{J}$'s are defined as
%$\mathcal{J}_{\nu+j-1}(\mu_i)\equiv (-1)^{j-1} K_{\nu+j-1}(\mu_i)$ 
%for $i=1,\cdots n$ and 
%$\mathcal{J}_{\nu+j-1}(\mu_i)\equiv I_{\nu+j-1}(\mu_i)$ 
%for $i=n+1,\cdots n+m$, 
%where $I_\nu$ and $K_\nu$ are the modified Bessel functions.
In this paper, we need the case with $(n, m)=(1, N+1)$ 
($N$ is the number of physical quarks):
%%%%%%%%%%%%%% Eq (n, m)=(1, N_f+1) partition function %%%%%%%%%%
\begin{eqnarray}
\label{eq:zero-mode1}
\mathcal{Z}^\nu_{1,1+N}(\mu_b |\mu_v ,\{\mu_s\})
&=&
\frac{1}{\prod_{s1=1}^{N}(\mu_{s1}^{2}-\mu_v^2)
\prod_{s2>s3}^{N}(\mu_{s2}^{2}-\mu_{s3}^{2})}
\nonumber\\
&&\hspace{-1in}
\times \det \left(
\begin{array}{ccccc}
K_\nu(\mu_{b}) & I_\nu(\mu_v) & I_\nu(\mu_{s1}) & I_\nu(\mu_{s2}) & \cdots\\
-\mu_{b} K_{\nu+1}(\mu_{b}) & \mu_v I_{\nu+1}(\mu_v) & \mu_{s1}I_{\nu+1}(\mu_{s1}) 
& \mu_{s2}I_{\nu+1}(\mu_{s2}) & \cdots\\
\mu_{b}^2 K_{\nu+2}(\mu_{b}) & \mu_v^2I_{\nu+2}(\mu_v) & 
\mu_{s1}^2I_{\nu+2}(\mu_{s1}) 
& \mu^2_{s2}I_{\nu+2}(\mu_{s2}) & \cdots\\
\cdots & \cdots & \cdots & \cdots & \cdots
\end{array}
\right),\nonumber\\
\end{eqnarray}
and $(n, m)=(2, N+2)$:
%%%%%%%%%%%%%%%% Eq (2,2+N_f) Z %%%%%%%%%%%%%%%%%%%%%%%%%%%
\begin{eqnarray}
\label{eq:zero-mode-2val}
\mathcal{Z}^\nu_{2,2+N}(\mu_{b 1},\mu_{b 2}|\mu_{v 1},\mu_{v 2},\{\mu_s\})
=\hspace{2.2in}
\nonumber\\
\frac{1}{(\mu_{b 2}^2-\mu_{b 1}^2)(\mu_{v 2}^2-\mu^2_{v 1})
\prod_{s1=1}^{N}
(\mu_{s1}^{ 2}-\mu_{v 2}^2)(\mu_{s 1}^{ 2}-\mu_{v 1}^2)
\prod_{s 2>s 3}^{N}(\mu_{s 2}^{ 2}-\mu_{s 3}^{ 2})}
\hspace{.3in}
\nonumber\\
\hspace{-.4in} 
\times \det \left(
\begin{array}{ccccc}
K_\nu(\mu_{b 1}) & K_\nu(\mu_{b 2}) & I_\nu(\mu_{v 1}) &I_\nu(\mu_{v 2}) 
%& I_\nu(\mu_1) %& I_\nu(\mu_2) 
& \cdots\\
-\mu_{b 1} K_{\nu+1}(\mu_{b 1}) &-\mu_{b 2} K_{\nu+1}(\mu_{b 2}) & 
\mu_{v 1}I_{\nu+1}(\mu_{v 1}) &
\mu_{v 2}I_{\nu+1}(\mu_{v 2})
%& \mu_1I_{\nu+1}(\mu_1) %& \mu_2I_{\nu+1}(\mu_2) 
& \cdots\\
\mu_{b 1}^2 K_{\nu+2}(\mu_{b 1}) &\mu_{b 2}^2 K_{\nu+2}(\mu_{b 2}) 
& \mu_{v 1}^2I_{\nu+2}(\mu_{v 1}) & \mu_{v 2}^2I_{\nu+2}(\mu_{v 2}) 
%& \mu_1^{s 2}I_{\nu+2}(\mu_1) 
%& \mu^2_2I_{\nu+2}(\mu_2) 
& \cdots\\
\cdots & \cdots & \cdots 
%& \cdots 
& \cdots
\end{array}
\right). \nonumber\\
\end{eqnarray}
Here $\mu_b=m_b\Sigma V$,  
$\mu_v=m_v\Sigma V$, where
$m_b$, $m_v$, denote the masses of the valence bosons, 
the valence quarks respectively.
%For the sea quark masses, $\mu_s = m_s\Sigma V$'s,
%$m_s$'s denote both of light masses $m_l$'s and heavier ones $m_h$'s.  
Partially quenched observables can be computed by differentiating
Eq.(\ref{eq:zero-mode1}) or (\ref{eq:zero-mode-2val})
with respect to suitable sources and subsequently 
taking the limit $\mu_b\to \mu_v$.

As building blocks, we use two quantities defined in 
Eqs. (\ref{eq:conden}) and (\ref{eq:Ddef}), 
%As building blocks, let us define two quantities,
%%%%%%%%%%%%%%%% Eq. building blocks
\begin{eqnarray}
\frac{\Sigma^{{\rm PQ}}_\nu(\mu_v,\{\mu_s\})}{\Sigma}
&\;\;\; \mbox{and }\;\;\; &
%&\equiv&
%\lim_{N\to 0} 
%\frac{1}{N}\langle\sum_v^{N}[U_0+U_0^\dagger]_{v v}
%\rangle_{U_0}
%\nonumber\\
%&=&
%-\lim_{\mu_b \to \mu_v}\frac{\partial}{\partial \mu_b} 
%\ln \mathcal{Z}^\nu_{1,1+N}(\mu_b | \mu_v, \{\mu_s\}),\\
%%%%%%%%%%%% Eq Delta Sigma definition %%%%%%%%%%%%%%%%%%%
%\begin{eqnarray}
%{\cal D}^{{\rm PQ}}_{\nu}(\mu_v,\{\mu_s\})
%&\equiv& \lim_{\mu_b\to \mu_v}\frac{
%\partial_{\mu_v} \partial_{\mu_b} 
%\mathcal{Z}^\nu_{1,1+N_l}(\mu_b|\mu_v,\{\mu_l\})}
%{\mathcal{Z}^\nu_{N_l}(\{\mu_s\})},\\
%\end{eqnarray}
%\end{eqnarray}
%and
%%%%%%%%%%% Eq D definition %%%%%%%%%%%%
%\begin{eqnarray}
D_\nu^{{\rm PQ}}(\mu_{v 1},\mu_{v 2},\{\mu_s\}).
%&\equiv& \hspace{2in}\nonumber\\
%&&\hspace{-1.2in}
%\lim_{\mu_{b 1}\to \mu_{v 1}, \mu_{b 2}\to \mu_{v 2}}
%\frac{
%\partial_{\mu_{v 1}}\partial_{\mu_{v 2}} 
%\mathcal{Z}^\nu_{2,2+N}(\mu_{b 1},\mu_{b 2}|\mu_{v 1},
%\mu_{v 2},\{\mu_s\})}{\mathcal{Z}^\nu_{N}(\{\mu_s\})}.
%%\lim_{\mu_{b 1}\to \mu_{v 1}, \mu_{b 2}\to \mu_{v 2}}
%\partial_{\mu_{v 1}}\partial_{\mu_{v 2}} 
%\mathcal{Z}^\nu_{2,2+N_f}(\mu_{b 1},\mu_{b 2}|\mu_{v 1},\mu_{v 2},\{\mu_i\}).
\end{eqnarray}
%We note that in the degenerate case we have the identity 
%%%%%%%%%%% Eq D \to Delta Sigma  %%%%%%%%%%%%%
%\begin{equation}
%D_\nu^{{\rm PQ}}(\mu_v,\mu_v,\{\mu_s\})=
%- {\cal D}^{{\rm PQ}}_{\nu}(\mu_v,\{\mu_s\}) .
%\end{equation}
%where the sea quark mass dependence is denoted by
%$\{\mu_s\} = \{ \mu_{1},....,\mu_{N}\}$. 
Note that in the degenerate limit $\mu_{v 1}=\mu_{v 2}=\mu_v$,
%%%%%%%%%%% Eq D property %%%%%%%%%%%%%%%%%%%%%%
\begin{eqnarray}
D_\nu^{{\rm PQ}}(\mu_{v},\mu_{v},\{\mu_s\})
&=&-\lim_{\mu_b \to \mu_v}\frac{\partial}{\partial \mu_b} \frac{\partial}{\partial \mu_v}
\frac{\mathcal{Z}^\nu_{1,1+N}(\mu_b | \mu_v, \{\mu_s\})}
{\mathcal{Z}^\nu_{0,N}(\{\mu_s\})}
\equiv -\frac{\Delta \Sigma^{{\rm PQ}}_{\nu}(\mu_v,\{\mu_s\})}{\Sigma}.
\nonumber\\
\end{eqnarray}
%for the integrals in eqs~(\ref{eq:zm_psi})-(\ref{eq:zm_psf}) and (\ref{eq:zm_vai})-(\ref{eq:zm_vaf}) and $\{\mu_s\}=\{\mu_l, \mu_h\}= \{\mu_{l_1},...,\mu_{l_{N_l}},\mu_{h_1},...,\mu_{h_{N_h}}\}$ in our second scheme where all zero-mode integrations are done exactly.

The needed formulae for one valence index are
%%%%%%%%%%%% 1-valence %%%%%%%%%%%%%%%%%
%%%%%%%%%%%%%%%%%%%%%%%%%%%%%%%%%%%%%%%%%
%%%%%%%%%%%%%%%%%%%%%%%%%%%%%%%%%%%%%%%%%
\begin{eqnarray}
%%%%%%%%%%%% Eq (U+U) %%%%%%%%%%%%%%%%% 
\frac{1}{2}\langle (U_{v v}+U^\dagger_{v v}) \rangle_U
&=&
\frac{\Sigma^{{\rm PQ}}_{\nu}(\mu_v,\{\mu_s\})}{\Sigma},\\
%%%%%%%%%%%% Eq (U+U)^2 %%%%%%%%%%%%%%%%% 
%\lim_{N_v\to 0}\frac{1}{N_v}\sum_v
\frac{1}{4}\langle 
(U_{v v}+U_{v v}^\dagger)^2 \rangle_U
&=&
\frac{\partial_{\mu_v}\Sigma^{{\rm PQ}}_{\nu}(\mu_v,\{\mu_s\})}{\Sigma}
-\frac{\Delta \Sigma^{{\rm PQ}}_{\nu}(\mu_v,\{\mu_s\})}{\Sigma},\\
%%%%%%%%%%%% Eq (U-U) %%%%%%%%%%%%%%%%% 
%\lim_{N_v\to 0}\frac{1}{N_v}\sum_v
\frac{1}{2}\langle (U_{v v}-U^\dagger_{v v}) \rangle_U
&=& -\frac{\nu}{\mu_v},\\
%%%%%%%%%%%% Eq (U-U)^2 %%%%%%%%%%%%%%%%% 
%\label{eq:diag-}
%\lim_{N_v\to 0}\frac{1}{N_v}\sum_v
\frac{1}{4}\langle 
(U_{v v}-U_{v v}^\dagger)^2 \rangle_U
&=& -\frac{\Sigma^{{\rm PQ}}_\nu(\mu_v,\{\mu_s\})}{\mu_v\Sigma} 
+ \frac{\nu^2}{\mu_v^2},
%\\
%%%%%%%%%%%% Eq (U^2-U^2) %%%%%%%%%%%%%%%%% 
%\frac{1}{4}\langle (U_{v v})^2-(U_{v v}^\dagger)^2 \rangle_U
%&=&\frac{\nu}{\mu_v^2}-\frac{\nu\Sigma^{{\rm PQ}}_\nu(\mu_v,\{\mu^s_i\})}{\mu_v\Sigma},
\end{eqnarray}
\begin{eqnarray}
%%%%%%%%%%%% Eq UU  %%%%%%%%%%%
\langle U_{v v}U^\dagger_{v v} \rangle_U
&=& \frac{1}{4}\langle 
(U_{v v}+U_{v v}^\dagger)^2 \rangle_U
- \frac{1}{4}\langle 
(U_{v v}-U_{v v}^\dagger)^2 \rangle_U\nonumber\\
&&\hspace{-0.6in}= \frac{\partial_{\mu_v}
\Sigma^{{\rm PQ}}_{\nu}(\mu_v,\{\mu_s\})}{\Sigma}
-\frac{\Delta \Sigma^{{\rm PQ}}_{\nu}(\mu_v,\{\mu_s\})}{\Sigma}
+\frac{\Sigma^{{\rm PQ}}_\nu(\mu_v,\{\mu_s\})}{\mu_v\Sigma} 
- \frac{\nu^2}{\mu_v^2}.
\end{eqnarray}
For two valence indices,
%%%%%%%%%%%% 2-valences %%%%%%%%%%%%%%%%%
%%%%%%%%%%%%%%%%%%%%%%%%%%%%%%%%%%%%%%%%%
%%%%%%%%%%%%%%%%%%%%%%%%%%%%%%%%%%%%%%%%%
\begin{eqnarray}
%%%%%%%%%%% Eq (U+U)11(U+U)22 %%%%%%%%%%%%
\frac{1}{4}\langle 
(U_{v v}+U_{v v}^\dagger) (U_{v^\prime  v^\prime }+
U_{v^\prime  v^\prime }^\dagger) \rangle_U
&=& D_\nu^{{\rm PQ}}(\mu_v,\mu_{v^\prime} ,\{\mu_s\}),\\
%%%%%%%%%%% Eq (U-U)11(U-U)22 %%%%%%%%%%%%
\frac{1}{4}\langle 
(U_{v v}-U_{v v}^\dagger) (U_{v^\prime  v^\prime }-U_{v^\prime  v^\prime }^\dagger)\rangle_U
&=& \frac{\nu^2 }{\mu_v\mu_{v^\prime} },\\
%%%%%%%%%%% Eq U11U22 %%%%%%%%%%%%
\langle U_{v v}U_{v^\prime  v^\prime }\rangle_U + 
\langle U^\dagger_{v v}U^\dagger_{v^\prime  v^\prime }\rangle_U
&=& 2D_\nu^{{\rm PQ}}(\mu_v,\mu_{v^\prime} ,\{\mu_s\})
+\frac{2\nu^2}{\mu_v\mu_{v^\prime} }.
\end{eqnarray}

Similarly,
\begin{eqnarray}
\label{eq:(U+U)^2}
%%%%%%%%%%% Eq (U+U)^2 = UU %%%%%%%%%%%%
\frac{1}{4}\langle 
(U_{v v^\prime }\pm U_{v^\prime  v}^\dagger)^2\rangle_U
&=& \frac{1}{4}\langle (U_{v^\prime  v}\pm U_{v v^\prime }^\dagger)^2\rangle_U
\nonumber\\
&=& \frac{\pm 1}{2}\langle U_{v v^\prime }U_{v^\prime  v}^\dagger\rangle_U
= \frac{\pm 1}{2}\langle U_{v^\prime  v}U_{v v^\prime }^\dagger\rangle_U
\nonumber\\
&=&
\frac{\pm 1}{\mu_v^2-\mu_{v^\prime}^2}
\left(\mu_v\frac{\Sigma_{\nu}^{{\rm PQ}}(\mu_v,\{\mu_s\})}{\Sigma}
-\mu_{v^\prime} \frac{\Sigma_{\nu}^{{\rm PQ}}
(\mu_{v^\prime} ,\{\mu_s\})}{\Sigma}\right),
\nonumber\\
%\\
%%%%%%%%%%% Eq U^2 + U^2 =0  %%%%%%%%%%%%
\frac{1}{4}\langle 
U_{v v^\prime }^2+ (U_{v^\prime  v}^\dagger)^2\rangle_U&=&0,
\end{eqnarray}
as well as
\begin{eqnarray}
%%%%%%%%%%% Eq (U+U)12(U+U)21 %%%%%%%%%%%%
\frac{1}{4}\langle 
(U_{v v^\prime }\pm U_{v^\prime  v}^\dagger)(U_{v^\prime  v}\pm U_{v v^\prime }^\dagger)\rangle_U
=\hspace{1.5in}\nonumber\\ 
\frac{1}{\mu_v^2-\mu_{v^\prime}^2}
\left(\mu_{v^\prime} \frac{\Sigma_{\nu}^{{\rm PQ}}(\mu_v,\{\mu_s\})}{\Sigma}
-\mu_v\frac{\Sigma_{\nu}^{{\rm PQ}}(\mu_{v^\prime} ,\{\mu_s\})}{\Sigma}\right),\\
\langle U_{v v^\prime}U^\dagger_{v v^\prime}
+U_{v^\prime v}U^\dagger_{v^\prime v}
\rangle_U =0.
\end{eqnarray}
were also derived in ref. \cite{Damgaard:2007ep}.

\section{Some Ward-Takahashi Identities at fixed topology}
\label{app:WTI}
\setcounter{equation}{0}
%%%%%%%%%%%%%%%%%%%%%%%%%%%%%%%%%%%%%%%%%%%%%%%%%%%
%%%%%%%%%%%%%%%%%%%%%%%%%%%%%%%%%%%%%%%%%%%%%%%%%%%
%%%%%%%%%%%%%%%%%%%%%%%%%%%%%%%%%%%%%%%%%%%%%%%%%%%

For the computation of vector and axial vector correlation
functions we needed a set of zero-mode expectation values
involving three zero-mode fields $U$. 
%(here for simplicity denoted $U$). 
These can be reduced to known integrals by means of exact identities
on the group manifold of $U(N)$. Such relations 
correspond to Schwinger-Dyson equations on the group
manifold and encode, in physics terms,
Ward-Takahashi Identities (WTI) of spontaneous chiral
symmetry breaking
in a sector of fixed topological charge $\nu$.
The derivation below follows the method described in detail in
Appendix B of ref. \cite{Damgaard:2001js}.

Let $t^a$ denote generators of $U(N)$ in a chosen representation,
here the fundamental. In addition, let $\epsilon_a$ be infinitesimal
parameters.
We introduce left-handed differentiation $\nabla^a$ on the group by means
of
%%%%%%%%%%%%%%% Eq differentiation defininition %%%%%%%%%%%%%%%
\begin{eqnarray}
F(e^{i\epsilon_at^a}U) ~=~ F(U) + \epsilon_a\nabla^aF(U) + \ldots
\end{eqnarray}
The derivatives $\nabla^a$ give rise to a standard Leibniz rule, and
left-invariance of the Haar measure on $U(N)$ ensures that
%%%%%%%%%%%%%%% Eq Haar measure invariance %%%%%%%%%%%%%%%%%%%%
\begin{eqnarray}
\int dU~ \nabla^a F(U) ~=~ 0 ~.
\end{eqnarray}
Choosing different functions $F(U)$ this simple identity generates
an infinity of exact relations on the coset of symmetry breaking for
the zero-mode fields. For the present purposes we can choose, $e.g.$,
%%%%%%%%%%%%%%% Eq choice of F %%%%%%%%%%%%%%%%%%%%%%%%%%%%%%%%
\begin{eqnarray}
F(U) ~\equiv~ \mathrm{Tr}[M_1U]\mathrm{Tr}[U^{\dagger}M_2]P(U),
\end{eqnarray}
where $M_1$ and $M_2$ are arbitrary $N\times N$
matrices, and
the Boltzmann weight $P(U)$ is defined in the obvious way:
%%%%%%%%%%%%%%% Eq weight %%%%%%%%%%%%%%%%%%%%%%%%%%%%%%%%%%%%%
\begin{eqnarray}
P(U) ~\equiv~ (\det U)^\nu 
\exp\left[\frac{\Sigma V}{2}\mathrm{tr}(\mathcal{M} U+
U^\dagger \mathcal{M})\right] ~.
\end{eqnarray}

Different choices of the matrices $M_1$ and $M_2$ 
lead to identities that are useful
in connection with the vector and axial vector correlators. For
example, $(M_1)_{ij} = \delta_{iv^\prime}t^a_{v^\prime j}$ and $(M_2)_{ij}
= \delta_{iv }\delta_{v j}$ (and the similar choice with indices 
$v$  and $v^\prime$  swapped) gives, 
after use of the $U(N)$ completeness relation (with a sum over $a$) 
$(t^a)_{ij}(t^a)_{kl} ~=~ \frac{1}{2}\delta_{il}\delta_{jk}$,
%%%%%%%%% Eq U UMU 1 %%%%%%%%%%%%%%%%%%%%%%%%%%%%%%%%%%%%%%%%%%%%%%
\begin{eqnarray}
\label{eq:UUMU1}
\langle U^\dagger_{v v }(U \mathcal{M} U)_{v^\prime v^\prime }\rangle_U
=\left\langle 
\mathcal{M}_{v^\prime v^\prime }U^\dagger_{v v }-\frac{2}{\Sigma V}
(N+\nu)U^\dagger_{v v }U_{v^\prime v^\prime }\right\rangle_U,
\end{eqnarray}
where we used $\mathcal{M}^\dagger=\mathcal{M}$,
and the hermitian conjugate relation  
(Note $(\det U)^\nu = (\det U^\dagger)^{-\nu}$),
%%%%%%%%%%%%%%%%%%%%%%%%%%%%%%%%%%%%%%
%%%%%%%%% Eq U UMU 2,3,4
\begin{eqnarray}
\label{eq:UUMU2}
\langle U_{v v }(U^\dagger \mathcal{M} U^\dagger)_{v^\prime v^\prime }\rangle_U
=\left\langle 
\mathcal{M}_{v^\prime v^\prime }U_{v v }-\frac{2}{\Sigma V}
(N-\nu)U_{v v }U^\dagger_{v^\prime v^\prime }\right\rangle_U.
\label{eq:UUMU4}
\end{eqnarray}
%together with the corresponding identities with indices swapped.

Another choice, $(M_1)_{ij}=\delta_{iv }t^a_{v^\prime j}$ and
$(M_2)_{ij} = \delta_{iv }\delta_{v^\prime j}$ gives
%%%%%%%%% Eq U UMU 5,6,7,8
\begin{eqnarray}
\label{eq:UUMU5}
\langle U^\dagger_{v^\prime v }(U \mathcal{M} U)_{v^\prime v }\rangle_U
=-\frac{2}{\Sigma V}
(N+\nu)\left\langle U^\dagger_{v^\prime v }U_{v^\prime v }\right\rangle_U = 0.
\label{eq:UUMU8}
\end{eqnarray}
See the appendix \ref{app:Uint} for the last equality to zero.
The hermitian conjugate is also vanishes,
%%%%%%% Eq U UMU last %%%%%%%%%%%%%%%%%%%%%%%%%%%%%%%%%%%%%%%%%%%%%%
\begin{eqnarray}
\langle U_{v^\prime v }(U^\dagger \mathcal{M} U^\dagger)_{v^\prime v }\rangle_U
=0.
%-\frac{2}{\Sigma V}
%(N_f-\nu)\left\langle U_{v^\prime v }U^\dagger_{v^\prime v }\right\rangle_U.
\end{eqnarray}